\documentclass{egpubl}
\usepackage[T1]{fontenc}
\usepackage{dfadobe}  

\usepackage{microtype}
\usepackage{graphicx}
\usepackage{color}
\usepackage{xcolor}
\usepackage{subcaption}
\usepackage{float}
\usepackage{tabu}                      
\usepackage{booktabs}                  
\usepackage{multirow}
\usepackage{listings}
\usepackage{textcomp}
\usepackage{makecell}
\usepackage{colortbl}
\usepackage{tablefootnote}
\usepackage{soul}
\usepackage{xfrac}
\usepackage{amsmath}
\usepackage{amssymb}
\usepackage{siunitx}
\usepackage{tikz}
\usetikzlibrary{shadows.blur}
\usepackage{tabularx}
\usepackage[table]{xcolor}

\usetikzlibrary{fadings, positioning}

\biberVersion
\BibtexOrBiblatex
\usepackage[backend=biber,bibstyle=EG,citestyle=alphabetic,backref=true]{biblatex} 
\electronicVersion
\PrintedOrElectronic

\definecolor{codegreen}{rgb}{0,0.6,0}
\definecolor{codegray}{rgb}{0.5,0.5,0.5}
\definecolor{codepurple}{rgb}{0.58,0,0.82}
\definecolor{backcolour}{rgb}{0.95,0.95,0.92}

\lstdefinestyle{mystyle}{
    backgroundcolor=\color{backcolour},   
    commentstyle=\color{codegreen},
    keywordstyle=\color{magenta},
    numberstyle=\tiny\color{codegray},
    stringstyle=\color{codepurple},
    basicstyle=\ttfamily\small,
    breakatwhitespace=false,         
    breaklines=true,                 
    captionpos=b,                    
    keepspaces=true,                 
    numbers=left,                    
    numbersep=5pt,                  
    showspaces=false,                
    showstringspaces=false,
    showtabs=false,                  
    tabsize=2
}

\usepackage{egweblnk} 

\usepackage[backend=biber]{biblatex}
\title[CuRast]{CuRast: Cuda-Based Software Rasterization for Billions of Triangles}

\author[D. Fellner \& S. Behnke]
{\parbox{\textwidth}{\centering Markus Schütz, Lukas Lipp, Elias Kristmann, Michael Wimmer}
        \\
{\parbox{\textwidth}{\centering TU Wien\\
       }
}
}

\begin{document}

\teaser{
\vspace{-2em}
 \begin{tikzpicture}
  \node[blur shadow={shadow blur steps=10, shadow xshift=1pt, shadow yshift=-1pt},
        inner sep=0pt] {
    \includegraphics[width=0.9\textwidth]{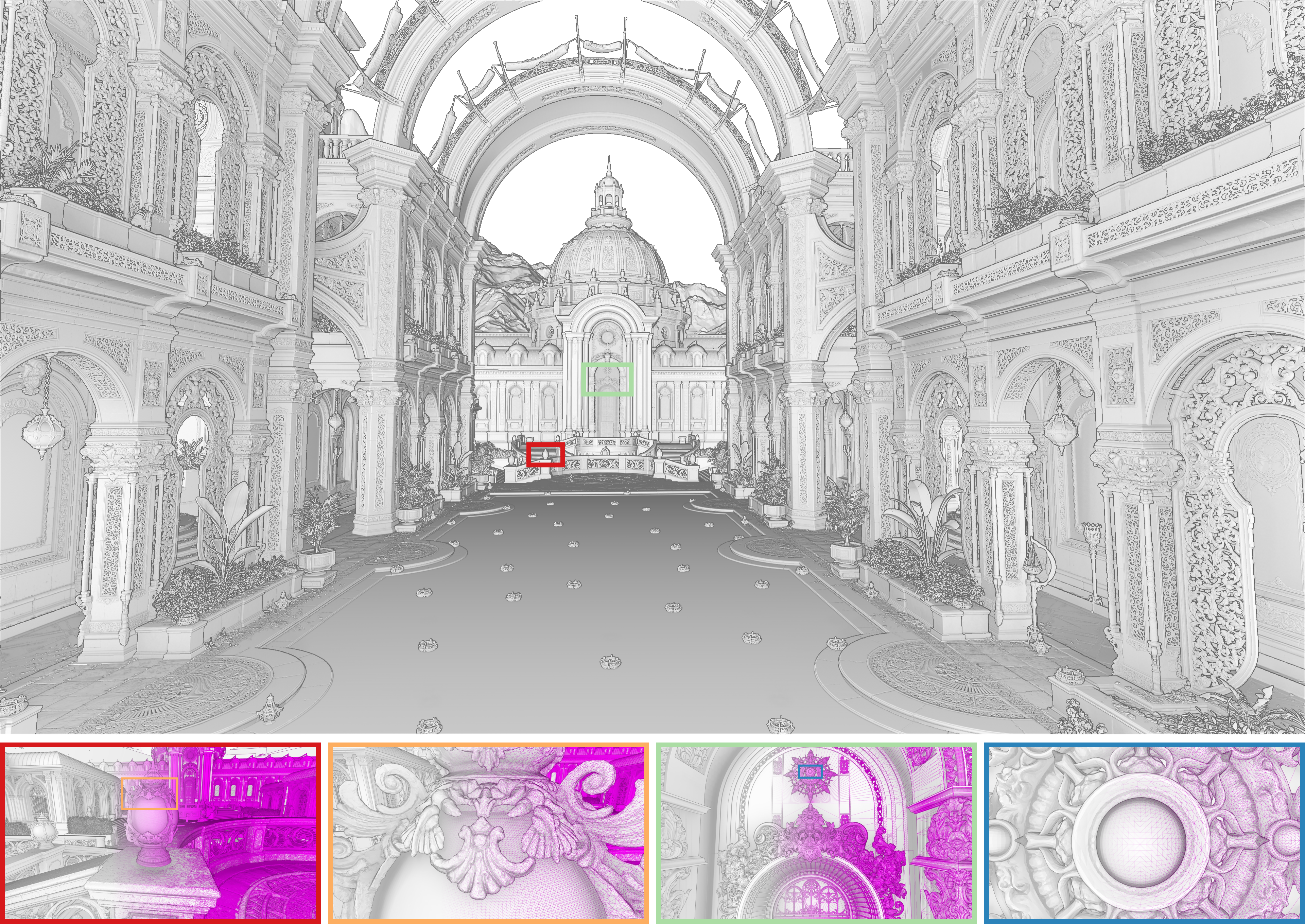}
  };
\end{tikzpicture}
 \centering
  \caption{Brute-force rendering the Zorah data set on an RTX 5090. 38.8GB of geometry loaded from an SSD, compressed to 21.7 GB, transferred to GPU and ready to render in 6.6 seconds. 18.9 billion triangles total; 13.5 billion triangles visible after frustum culling. Rendered in 67.3 milliseconds into a 3840$\times$2160 framebuffer with screen space ambient occlusion and eye-dome lighting enabled. 
  }
\label{fig:teaser}
}

\maketitle

\begin{abstract}
Previous work shows that small triangles can be rasterized efficiently with compute shaders. Building on this insight, we explore how far this can be pushed for massive triangle datasets without the need to construct acceleration structures in advance.

\vspace{0.3\baselineskip}
\textbf{Method:} A 3-stage rasterization pipeline first rasterizes small triangles directly in stage 1, using atomicMin to store the closest fragments. Larger triangles are forwarded to stages 2 and 3. 

\vspace{0.3\baselineskip}
\textbf{Results:} CuRast can render models with hundreds of millions of triangles up to 2-5x (unique) or up to 12x (instanced) faster than Vulkan. Vulkan remains an order of magnitude faster for low-poly meshes. 

\vspace{0.3\baselineskip}
\textbf{Limitations:} We currently focus on dense, opaque meshes that you would typically obtain from photogrammetry/3D reconstruction. Blending/Transparency is not yet supported, and scenes with thousands of low-poly meshes are not implemented efficiently.

\vspace{0.3\baselineskip}
\textbf{Future Work:} To make it suitable for games and a wider range of use cases, future work will need to (1) optimize handling of scenes with tens of thousands of nodes/meshes, (2) add support for hierarchical clustered LODs such as those produced by Meshoptimizer, (3) add support for transparency, likely in its own stage so as to keep opaque rasterization untouched and fast.

\vspace{0.3\baselineskip}
\textbf{Source Code:} https://github.com/m-schuetz/CuRast

\end{abstract}  








\section{Introduction}

Dedicated hardware rendering pipelines have been the dominant approach for visualizing three-dimensional triangle meshes for decades. However, prior work on micropolygon and triangle cluster rendering, such as Nanite~\cite{karis2021nanite}, have demonstrated that hardware rasterization can be outperformed in certain scenarios, particularly when rendering highly detailed geometry composed of pixel-sized triangles. On average, Nanite's software rasterizer achieves roughly three times the throughput of the hardware pipeline, with even larger gains for pure micropolygon geometry. For larger triangles or triangle clusters, Nanite falls back to traditional hardware rasterization.

In this paper, we further explore the potential of GPGPU-accelerated software rasterization and investigate the limits achievable through brute-force processing of unstructured triangle meshes. By \emph{unstructured}, we mean that no spatial acceleration structures or hierarchical levels of detail (LOD) are precomputed. We deliberately avoid such preprocessing for two reasons: first, constructing these structures requires a costly preprocessing step, and second, they typically must be recomputed whenever the underlying mesh is modified. 

GPU-based software rasterizers typically also aim to support a wide range of features, including blending and support for transparent geometry. These features can be a defining aspect in the architecture of a rasterization pipeline and potentially slow down the processing of opaque geometry. In this work, we omit blending and focus on maximizing performance for opaque geometry. Transparent geometry could, in theory, be implemented in subsequent passes that execute after our pipeline. 

Our contributions to the state of the art are as follows:

\begin{itemize}
    \item A software rasterizer capable of rendering up to a billion (unique geometry) to 4 billion (instanced geometry) of triangles in real time, without the need to create spatial acceleration structures or hierarchical levels of detail in advance.
    \item A 3-stage pipeline that is highly optimized for small, but also handles medium and large triangles.
    \item Adapting visibility buffer indexing for an arbitrary mix of nodes/instances/triangles, rather than supporting a fixed number of nodes and a fixed maximum of triangles per node. 
\end{itemize}

To manage expectations about the scope of this work, following limitations and observations apply:

\begin{itemize}
    \item We specifically focus on dense and opaque geometry. Blending is not supported.
    \item Our implementation currently focuses on triangle-dense meshes and scales poorly with tens of thousands of low-density meshes.
    \item Likewise, instancing is targeted towards meshes with several hundreds of thousands to millions of triangles.
    \item Larger triangles are handled in a "good enough" fashion with glaring potential for improvement.
\end{itemize}

Given these benefits and limitations, we believe the results of this work are particularly relevant for applications involving dense and editable or animated meshes, as our approach does not rely on acceleration structures that require recomputation after modifications. Representative use cases include photogrammetry reconstruction and processing pipelines that generate and curate models containing tens of millions to billions of triangles. Additional examples include content creation tools such as Blender, where users perform arbitrary editing operations on complex geometry. It is not yet suitable for games as these work largely with non-editable geometry that benefits from pre-constructed LODs, and because they typically comprise scenes with tens of thousands of low-density meshes, which our implementation is currently not made for. 

\section{Related Work}

\subsection{Software Rasterization}

Software rasterization has a long history that predates dedicated GPU hardware, and it continues to be of interest as modern GPUs provide powerful and fast compute pipelines for custom rendering algorithms. One of the most fundamental concepts of efficient rasterization pipelines -- the depth buffer -- dates all the way back to 1974, long before GPUs became commonplace~\cite{catmull1974subdivision}. Since GPUs became widely available, software rasterization was primarily motivated by education and research, but in the recent decade the interest has shifted towards targeting specialized use cases in which custom programs can outperform the more generalized hardware pipeline that has to support a wide range of scenarios. 

\subsection{Triangles}

FreePipe~\cite{liu2010freepipe} introduces the concept of atomic-min to efficiently store the closest fragment inside pixels without the need for inter-thread communication and sorting stages. Since CUDA was limited to 32 bit atomics back then, which did not allow atomically writing depth and color values with a single atomic operation, they suggested performing two 32 bit atomic-min operations with the same 20 bit depth value but different 12 bits of the color value into the same pixel that is separated into two buffers. Afterwards, they extract the 12 bit parts located in different buffers, and fuse them back into a 24 bit color value. Brunhaver et al.~\cite{brunhaver2010hardware} design a hardware system targeted towards micropolygons that is able to process triangles at the rate of an GTX 480, but with less than 1\% of the die size and power usage. CudaRaster~\cite{10.1145/2018323.2018337} employs a hierarchical approach where triangles are first queued into bins, then tiles, and finally rasterized in pixels. They also structure their pipeline to honor the order of triangles during rendering via a sort-middle~\cite{molnar1994sorting} approach, enabling order-dependent algorithms such as front-to-back blending operations of transparent triangles. Weber built a software rasterization approach for micropolygons, where they first lock a pixel sample, update the color value, then unlock it~\cite{WEBER-2015-PRA1}. Kenzel et al.~\cite{Kenzel:2018:CURE} propose cuRE, a streaming architecture that performs multiple rasterization stages in parallel rather than one after another. This approach avoids amassing large workloads in queues between stages, as the queues are quickly processed in a timely manner. They also aim for a rich set of features comparable to DirectX 9, and evaluate with a large set of more than 100 test scenes captured from video games, with about 0.3 to 8 million triangles. cuRE compares to Piko~\cite{Piko}, FreePipe, CUDARaster and OpenGL, and finds that CUDARaster generally performs fastest within a factor of 4-6 to OpenGL, followed by cuRE. cuRE, on the other hand, scales better with multiple draw calls and higher screen resolutions. FreePipe is slower than either for most scenes due to utilizing only one thread per triangle, which becomes an issue for scenes with large triangles. 

\subsection{Points and Splats}

Software rasterization has seen particular success in the field of point-based rendering, which poses fewer challenges as points are typically always small, often pixel-sized. Due to this, we rarely run into load-balancing issues caused by processing different-sized primitives (with the notable exception of Gaussian Splats). Günther et al.~\cite{Gnther2013AGP} proposed a compute-based rendering pipeline based on spin loops: Each thread trying to draw a point to a pixel first attempts to lock it with atomic-CAS, then updates and subsequently unlocks it. Marrs et al.~\cite{Marrs2018Shadows} use 32 bit atomic-min operations to efficiently construct multiple depth maps. Schütz et al.~\cite{SCHUETZ-2021-PCC} use 64 bit atomic-min operations to efficiently write interleaved depth and color values into a framebuffer, then extract the color values for display on screen. Neural point-based renderers ADOP~\cite{ruckert2022adop} and NePO~\cite{NePO_NeuralPointOctrees} use atomic-min-based rasterization to quickly construct multi-resolution framebuffers that are then fed to a neural renderer that fills holes and shades the final image. One of the biggest success stories of software rasterization and point-based rendering might be 3D Gaussian Splatting (3DGS)~\cite{kerbl3Dgaussians}. 3DGS proposes using translucent gaussian primitives for scene reconstruction from photos, resulting in high-quality models of the real world. The 3-dimensional gaussians are then rendered by approximating them through 2-dimensional gaussians in screen space, sorting them by depth, and blending them from front to back. To efficiently process splats with wildly different sizes, they are partitioned into 16$\times$16 pixel tiles, and they then rasterize all splats in a tile with one CUDA thread block. Since then, a plethora of research has proposed optimizations to various aspects of the training and rasterization pipeline. We refer to Hahlbohm et al.~\cite{hahlbohm2026fastergs} as a recent summary of a variety of these optimizations.

Besides 3DGS, Dreams (PS4)~\cite{2015learning} and Nanite~\cite{karis2021nanite} are two notable software rasterizers that demonstrate the practical use of custom rendering pipelines. The developers of Dreams discovered that they could rasterize small splats faster with a custom atomics-based renderer than the standard PS4 hardware pipeline. Nanite is primarily an LOD system for massive meshes that renders clusters of 128 triangles with a desired level of detail. However, they discovered that rasterizing these fine-grained clusters with near pixel-sized triangles is often faster through custom compute shaders, than through the hardware rasterizer. For large triangles, they fall back to hardware rasterization. 

\subsection{Visibility Buffers}

Visibility buffers were initially introduced under the term \emph{item buffers} in the context of ray tracing~\cite{10.1145/357332.357335}. For each pixel, they store the id of the element that corresponds to the first hit of a ray. They were later re-introduced and formalized for triangle-primitives under the term \emph{visibility buffer}, as an alternative to deferred shading~\cite{VisibilityBuffer}. Instead of constructing feature-rich G-Buffers, visibility buffers store triangle and model/instance indices of the visible triangle for each pixel, which simplifies the rasterization stage and allows us to look up all necessary data in the shading stage. Alternatively, visibility buffers could also store references to shading values~\cite{VisibilityBuffer2}, but in the context of this paper, we assume they reference individual triangles. Visibility buffers have gained popularity for software rasterization as they allow us to address all attributes of a visible triangle, despite the limited amount of information we can write to a framebuffer with 64 bit atomic-min-based rasterizers. 

\subsection{Geometry Optimization}

The order of vertices and indices has a substantial impact on rendering performance due to factors like access to coalesced data in memory and caching, as well as potential vertex reuse mechanisms in GPUs. Kenzel et al.~\cite{Kenzel:2018:OVR} and Kerbl et al.~\cite{RevisitingVertexCache} studied the behaviour of vertex caching and reuse on various modern GPUs, and propose strategies to optimize the meshes for better reuse, as well as strategies to implement reuse in software rasterization pipelines. Schütz et al.~\cite{SCHUETZ-2021-PCC} rearrange point clouds such that some spatially close points are also close in memory to promote coalesced memory access, but avoid too much locality that leads to contention when rendering thousands of overlapping points to the same pixel. Bene and Valasek~\cite{HelperLaneOptimization} investigate various triangulation strategies for polygons to find those that render the fastest. 

Meshoptimizer is an open source project that implements strategies that promote locality and vertex reuse~\cite{meshoptimizer}. In this paper, we use it to obtain fair comparisons with a Vulkan-based renderer that greatly benefits from it, and also observe its impact on our own software rasterization pipeline.

\section{Preliminaries}

In this section, we recap important basics/prior work we build on in more detail, particularly atomic-min-based software rasterization and visibility buffer rendering pipelines. 

\subsection{Atomics-Based Depth Testing}

In 2010, FreePipe~\cite{liu2010freepipe} proposed using atomic min operations to efficiently write the closest depth value to each pixel in a massively parallel rasterizer. Because global atomics implicitly synchronize concurrent writes, this approach eliminates the need for inter-thread communication or fragment queuing and sorting. At the time, however, the method was limited to 32-bit atomics, which made it difficult to attach information in addition to the mandatory depth value. With the introduction of 64-bit atomic min/max operations, it became possible to attach additional payload such as color values or primitive IDs, enabling atomic updates to an interleaved depth+payload framebuffer:

\begin{lstlisting}[language=Java,label={lst:basicRasterization},captionpos=b]
u32 udepth = __float_as_uint(depth);
u32 payload = ...; // color or primitive ID
u64 fragment = u64(udepth) << 32 | u64(payload);
atomicMin(framebuffer[pixelID], fragment);
\end{lstlisting}

The PS4 game "Dreams" utilized that additional payload to create a highly efficient brush-stroke/splat based rasterizer that was already able to outperform the hardware rasterization pipeline of the PlayStation 4~\cite{2015learning}. While 64 bit atomics are fast, the still fairly small payload is not suitable for rendering algorithms that rely on multiple render targets, such as deferred rendering with rich G-Buffers. To work around this limitation, we can either (A) compute each fragment's final shaded color value directly during triangle rasterization, which is prohibitively expensive, or (B) we can store primitive indices instead to access all of the visible triangle's attributes in a deferred resolve pass. The second approach is what is commonly referred to as "visibility buffers" and popularized by Nanite~\cite{karis2021nanite}.

\subsection{Visibility Buffers}

Visibility buffer rendering and deferred shading with G-Buffers are closely related. Both approaches aim to rasterize all geometry first, and defer the shading to a full-screen resolve pass in order to apply expensive shading operations only to visible fragments. The difference is that deferred rendering creates multiple (or packed) render targets that contain attributes that are necessary for shading (e.g., albedo, normals, material id/flags, roughness, etc.), while visibility buffers only store the ID of the visible mesh and triangle in each pixel. With these IDs, we are then able to fetch all of the data that we need for shading in the full-screen resolve pass.

\begin{figure}[H]
    \centering
    \includegraphics[width=\columnwidth]{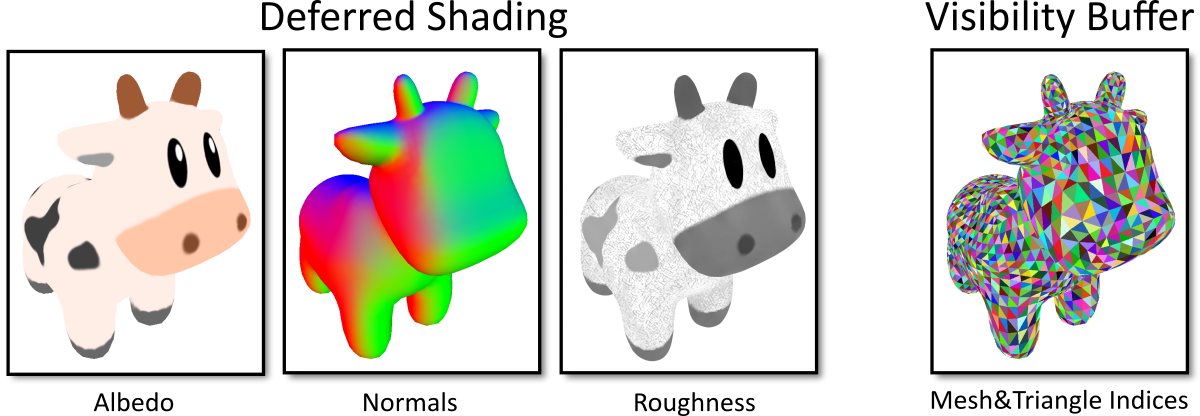}
    \caption{Deferred shading creates G-Buffers with various attributes that may be needed for the shading pass. Visibility buffers only store triangle indices.}
    \label{fig:hqs_artifacts}
\end{figure}

Nanite~\cite{karis2021nanite} demonstrated the efficiency and viability of software-rasterization with visibility buffers in a practical and widely deployed gaming engine. It assigns 30 bit for depth values and a payload of 34 bit for the triangle ID. The triangle ID itself is composed of 27 bit for the cluster index and 7 bit to address the triangle within the cluster. With this assignment of bits, they are able to address 134 million clusters and 128 triangles per cluster. 

A major advantage of visibility buffers is that they significantly reduce memory bandwidth pressure during geometry rasterization, since we only need to load vertex indices and positions. Other attributes, such as uv-coordinates, normals, textures, etc., are not needed, unless they affect vertex positions (e.g. bump maps or height maps). This further accelerates the geometry processing stage compared to deferred rendering, which still fetches data for numerous fragments that will not be visible in the final image. Another advantage of visibility buffers is that we get object picking for free. 

A disadvantage of visibility buffers is that the resolve pass is significantly more expensive: We now need to load the triangle geometry, project it again, compute interpolated uv coordinates, etc. Visibility buffers may also suffer from caching issues because if triangles are roughly pixel-sized, adjacent pixels load vertex data from different and quasi-random regions in memory~\cite{VisBufferMatGraphs}. Another disadvantage of visibility buffers is that computing the mip map level can be tricky for large triangles that intersect with the near plane. If one or two vertices are behind the near plane, we can not inter- or extrapolate uv-coordinates in adjacent pixels in screen space, which would allow us to cheaply compute the pixel's footprint in texture space. We therefore shade triangles in world space (see Section~\ref{sec:visibilitybuffer_implementation}).

\section{Method}

Our method draws massive triangle models entirely in CUDA and supports following functionality: Indexed meshes, instancing, textures or vertex colors, mip mapping, and massive amounts of small but also large triangles. 
We furthermore experiment with compressed/quantized indices between 8 to 32 bit and 16-bit fixed-precision coordinates. 

\begin{figure}[htp]
    \centering
    \includegraphics[width=\columnwidth]{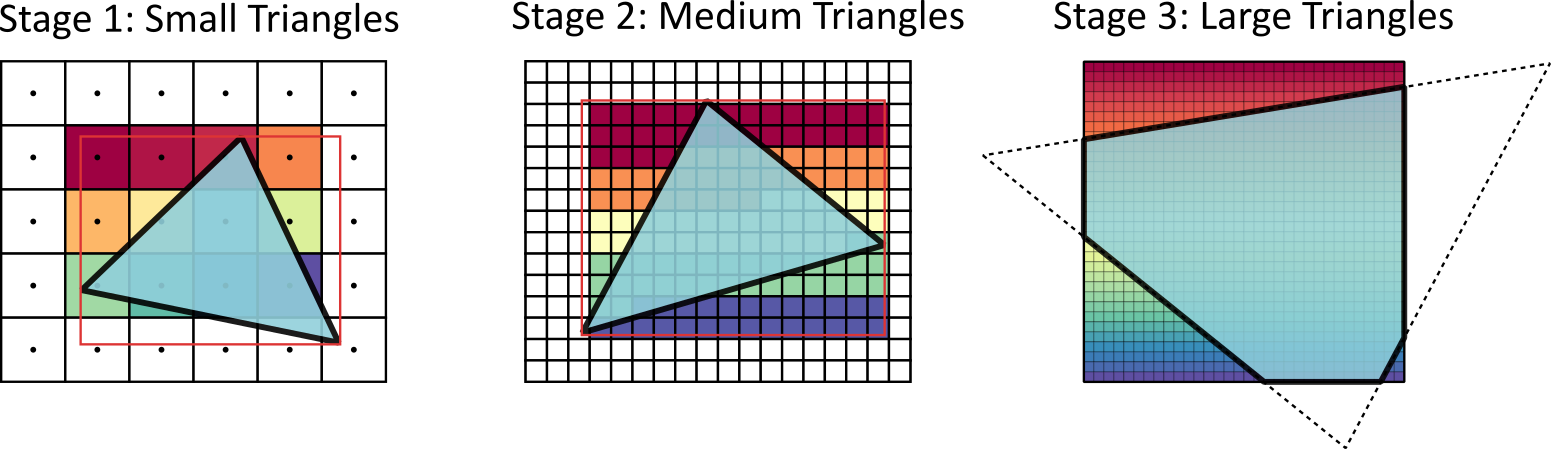}
    \caption{Rasterization Stages. Pixels colored by iteration counter. \textbf{Stage 1}: A single thread iterates over all sample positions inside the triangle's bounding box. \textbf{Stage 2}: A warp (32 threads) iterates over all samples inside the bounding box. \textbf{Stage 3}: A workgroup (64 threads) iterates over all samples in a 64x64 tile to rasterize a portion of the triangle.}
    \label{fig:hqs_artifacts}
\end{figure}

As we primarily focus on massive models with hundreds of millions to billions of small triangles, our method first launches a CUDA kernel with one thread per triangle, under the assumption that triangles can easily be rasterized by a single thread, and multiple threads in a warp have balanced workload as they process similar-sized triangles. This assumption will, of course, be frequently violated so we implement two additional stages that handle medium-sized and large triangles:

\begin{itemize}
    \item \textbf{Stage 1}: Launch 1 thread per triangle, rasterizing it directly if it is small, or adding it to a global queue for the next stage if it is either too large or intersects the near plane.
    \item \textbf{Stage 2}: Launch a warp (32 threads) per triangle that was queued by stage 1, rasterizing it directly if it is medium-sized, or splitting it into parts and adding each part to a queue if it is either very large or intersects the near plane.
    \item \textbf{Stage 3}: Launch 64 threads for each part of a triangle that was queued by stage 2.  
\end{itemize}

\emph{Small} triangles were experimentally determined as those whose screen-space bounding box covers less than 128 pixels, and \emph{medium} triangles as those covering less than 4096 pixels.

On the host side, before launching the rasterization stages, we first perform frustum culling and assemble a list of meshes, each comprising transformation matrices and pointers to vertex and index buffers. Crucially, we also compute a cumulative triangle count for each mesh and instance (i.e. the exclusive prefix sum), which we will use to store the absolute triangle ID over all visible meshes in the visibility buffer. This data is then copied to the GPU, and we are ready to launch the 3 stages of our rasterization kernels.

Experienced readers will notice that we perform naïve boundary constraints by processing all fragments inside the triangle's bounding box, despite well-known approaches that only traverse fragments within the triangle~\cite{10.1145/54852.378457}. We initially tried to do so but found that the additional effort that was required to avoid wasted work is often more expensive than unnecessary but cheap work, especially for massive, dense data sets that produce pixel-sized triangles.

\subsection{Stage 1: Rasterizing Small Triangles}

The premise of this stage is that each triangle is small enough such that a single thread can render it in a timely manner. If they are too large, we instead put them in a global queue for the next stage. 

\textbf{Looping Through Batches of 256 Triangles}: We perform a persistent-kernel launch using cooperative groups since we are rendering numerous meshes with varying numbers of triangles, and therefore cannot easily map a global thread index to the corresponding triangle. A limited set of workgroups with 256 threads keeps looping until all triangles are rendered. Each specific workgroup obtains the next set of triangles it renders by atomically incrementing a global work counter: \emph{atomicAdd(numTrianglesRendered, 256)}. If the returned triangle index is outside of the current mesh, the workgroup advances through the list of meshes until it encounters the mesh that contains the corresponding triangles.

\textbf{Preparing a Batch of 256 Triangles for Rasterization}: Each thread loads the vertex data of its triangle and transforms it to view space. Unlike classical graphics APIs, we do not use a projection matrix. Instead, we perform perspective projection through an element-wise multiplication of the view-space coordinates with following projection-vector:

\begin{equation}
f = \frac{1}{\tan\left(\frac{\text{fovy}}{2}\right)}
\end{equation}
\begin{equation}
\mathbf{P} = \begin{bmatrix}
\frac{f}{\text{aspect}} & f & -1
\end{bmatrix}
\end{equation}

Afterwards, the x and y components are divided by the depth (z component). The result is a normalized-device-coordinate with x and y between -1 and 1, and z storing the positive, linear and non-normalized depth value, starting from the camera position.

We continue to perform frustum culling, discarding/skipping any triangle that is entirely outside the visible volume. For the remaining triangles, we compute the screen-space bounding box, or mark it as \emph{nontrivial} if it intersects the near plane. The latter is done because triangles that intersect the near plane require clipping to obtain accurate screen-space bounding boxes, but we want to keep this stage as simple as possible and avoid increased register pressure and expensive branches, which would slow down the processing of nicely behaved triangles. To optimize for scenarios with massive amounts of micro-polys, we also discard/skip triangles whose bounding box does not intersect with the center of any pixel, followed by performing backface culling. Finally, we put all triangles that either intersect with the near plane or whose bounding box covers more than 128 pixels into the queue for the next stage. The remaining triangles are then rasterized.

\textbf{Rasterization}: Rasterization follows a straight-forward bounding-box-rasterization scheme. Each thread loops over the y (outer loop) and x (inner loop) coordinates inside the bounding box, computes the barycentric coordinates of the triangle, and checks if the barycentric coordinates lie within the triangle boundary. If the fragment lies inside the triangle, we continue to perform a 64-bit atomicMin to update the framebuffer's pixel with a combination of depth (most significant bits) and global triangle index (least significant bits). To support scenes with tens of billions of triangles, we assign 28 bit to the depth value, and 36 bit to the global triangle ID. Since depth values are always positive, one bit is taken from the sign, and three bits are taken away from the mantissa. As the 28 bit depth value is stored in the most significant bits, the fragment with the smallest depth will prevail in the resulting framebuffer. The encoded global triangle index is later used in a resolve/screen pass in order to find and process the triangle that occupies that pixel. 

For efficient perspective-correct interpolation of depth values, we precompute the inverse depth $\frac{1}{z}$ for each vertex outside of the loop, then interpolate them according to the barycentric coordinates $s,t,v$ and a fast division intrinsic inside the loop:

\begin{lstlisting}[language=Java,label={lst:basicRasterization},captionpos=b]
float depthI = v * z0I + s * z1I + t * z2I;
float depth = __fdividef(1.0f, depthI); 
\end{lstlisting}

Of note is that \_\_fdividef is not just a faster approximate division; it also significantly reduces the kernel's register usage from 55 to 48, and therefore improves its occupancy.

For barycentric coordinates, we precompute their changes across the x and y directions outside of the loop, and then increment them as we loop over the pixels of the bounding box.

\subsection{Stage 2: Rasterizing Medium Triangles}

We now launch 32 threads per triangle that was queued by stage 1, so that we can process these more demanding ones with additional compute power. For any triangle whose bounding box covers up to 4096 pixels, the 32 threads perform a 1-dimensional loop over the number of pixels, incrementing the loop counter by 32 each iteration. The 1D index is then mapped to the 2D coordinate within the bounding box via $x = i \bmod\text{width}$ and $y = \frac{i}{\text{width}}$, but otherwise the rasterization logic remains the same as in stage 1. 

A major difference from the first stage is that we now also perform an accurate and expensive screen-space bounding box calculation for each triangle by clipping them via the Sutherland-Hodgman algorithm. However, we do not actually split triangles that intersect the frustum into multiple smaller ones. Our intention is to (1) identify and measure huge triangles (which often intersect the near plane), so that we can partition them into a suitable number of tiles for stage 3, and (2) to obtain the bounding box of the visible part of the triangle, which may be a fraction of the bounding box of the entire triangle. 

Triangles that are larger than 4096 pixels are split into smaller parts according to 64$\times$64 pixel tile boundaries. For each part, we add one entry into a global queue comprising the triangle ID, mesh ID, and the tile coordinate.

\subsection{Stage 3: Rasterizing Large Triangles}

For each part of a triangle that was queued in stage 2, we launch one workgroup comprising 64 threads. Each workgroup then processes the pixels inside the 64 x 64 tile line-by-line. For each processed pixel, the corresponding thread performs a ray cast in world space in order to determine whether the triangle is hit, and at which location. Depth and total triangle index are then stored inside the pixel using a 64-bit atomicMin.

The reason we render triangles in this stage with ray-triangle intersections in world space is that the barycentric-coordinate-approach in normalized device coordinates would require us to clip triangles that intersect the near plane. 

\subsection{Screen-Space Resolve/Shading Pass}
\label{sec:visibilitybuffer_implementation}

In this pass, we texture and shade the triangles whose global indices are stored in the pixels of the visibility buffer. We launch a 2-dimensional CUDA kernel with one thread per pixel. For each pixel, we decode depth (28 bit) and global triangle index (36 bit). 

One of the challenges we faced was how these 36 bit of the visibility buffer would be able to address individual triangles in scenes with tens of billions of triangles, tens of thousands of meshes, and up to tens of millions of triangles per mesh. Traditionally, visibility buffers assign a fixed number of bits to mesh indices and triangle indices, so we can either increase the number of supported meshes, or the number of triangles per mesh. This was not feasible for Zorah, which required much of both. So instead of storing a combination of mesh and triangle index inside the visibility buffer, we store the cumulative triangle index and now perform a binary search inside the resolve pass to find the mesh that this triangle belongs to. A 64-bit integer array storing the exclusive prefix sum of triangle counts accelerates the search for the correct mesh's index.

\begin{figure}[htp]
    \centering
    \includegraphics[width=0.5\columnwidth]{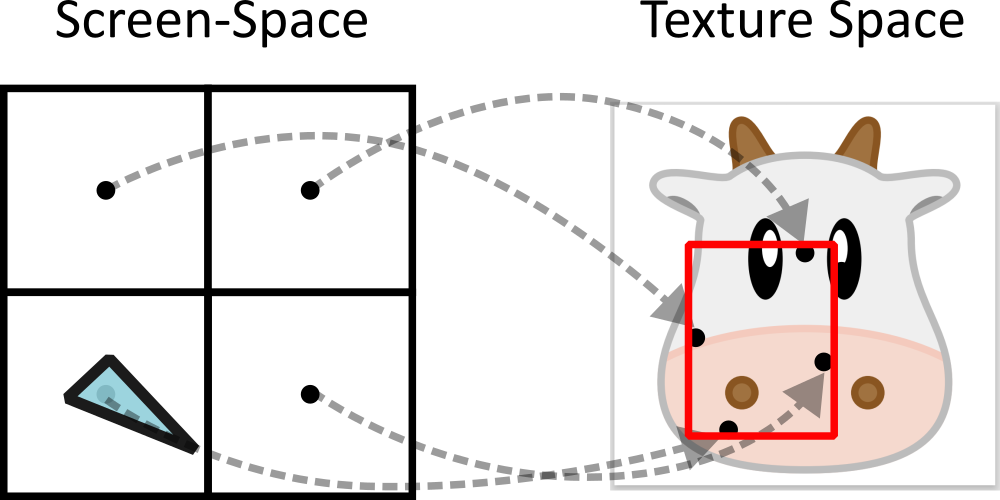}
    \caption{Finding an appropriate mip map level by intersecting the triangle's plane at the current and three adjacent pixels, extrapolating uv coordinates, and computing the extent in texture space.}
    \label{fig:mipmapping}
\end{figure}

\textbf{Shading and Mip Mapping:} Having identified the mesh that corresponds to the global triangle index, we proceed to load the triangle’s vertex data and perform shading in world space. This choice is primarily motivated by the need to correctly determine the mipmap level for large triangles intersecting the near plane, for which we can not easily compute deltas to adjacent pixels in screen-space. Estimating a suitable mipmap level requires computing the pixel footprint in texture space. To achieve this, we cast rays from the camera through the current pixel as well as its right, top, and top-right pixels, and intersect these rays with the triangle’s plane, as shown in Figure~\ref{fig:mipmapping}.

Using the resulting world-space intersection points together with the triangle’s vertex positions, we compute barycentric coordinates for each intersection. These are then used to interpolate or extrapolate the corresponding UV coordinates. Note that some intersections, and thus their UV coordinates, may lie outside the triangle, which is acceptable for mipmap level estimation. Once the UV coordinates of all four samples are obtained, we compute their extent (delta) in texture space and derive the appropriate mipmap level.

\subsection{Instancing}
\label{sec:compression}

Memory bandwidth is one of the major bottlenecks during rasterization, particularly when rendering massive amounts of small triangles that require fewer compute resources. Instancing allows us to reduce this bandwidth bottleneck because instead of loading and rasterizing a set of triangles several times, we only need to load it once to rasterize it several times. 

We adapt stage 1 such that each workgroup loads indices and position of a triangle once, then loops through all instances and draws the triangle with the corresponding transformation matrices. Stages 2 and 3 remain unchanged because most instances of a triangle are already likely to be finished by stage 1, so the kernel-overhead of handling instancing does not benefit the later stages. Unfortunately, this overhead does affect the stage 1 kernel, i.e., it increases the amount of required registers and thus reduces occupancy. We therefore suggest compiling and using two variations of the stage 1 kernel, one without and one with instancing. 

\subsection{Compression}

Massive data sets such as Zorah~\cite{Zorah,ZorahGDC2025} do not fit into GPU memory and thus need to be streamed or compressed. In order to fit the original 38.8GB data set into the 24GB VRAM of an RTX 4090, we perform following two compression schemes.

\textbf{Compressing Indices}: While loading the indices of a mesh, we compute the minimum and maximum index and the resulting $\mathit{span} = \max - \min$. The number of bits required to store indices in that span is given by the ceiling of the base-2 logarithm:
\begin{equation}
\mathit{bitsPerIndex} = \lceil \log_2(\mathit{span}) \rceil
\end{equation}
For example, if an index buffer comprises index values between 2500 and 3000, we must represent 500 distinct values, which can be encoded in 9 bits.

\textbf{Compressing Positions}: For positions, we encode them as 16-bit fixed-precision integers. 16 bits afford $65\,536$ distinct values, meaning a mesh spanning 65 meters retains a vertex precision of 1\,mm. The conversion to fixed-precision coordinates relative to the mesh's bounding box is:
\begin{lstlisting}
uint16_t X = 65536.0 * (pos.x - min.x) / size.x;
\end{lstlisting}

With these two lightweight compression schemes, and by dropping the 2\,GB of UV coordinates which only exist for a fraction of meshes, we reduced the Zorah data set from 38.8\,GB to 21.7\,GB, sufficiently small to fit on an RTX 4090. Similar to instancing, we found that compiling separate CUDA kernels for uncompressed and compressed data slightly improves performance for the uncompressed case. The presence of a branch inside the kernel that checks if data is compressed would otherwise add up to 1-2\% to the runtime in scenes with billions of triangles.

\section{Evaluation}

\newcommand{\screenshotsize}{0.35}
\newcommand{\scenerow}[9]{%
  \begin{minipage}[t]{\screenshotsize\textwidth}\vspace{0pt}
    \includegraphics[width=\textwidth]{#1}
  \end{minipage} &
   \begin{minipage}[t]{\screenshotsize\textwidth}\vspace{0pt}
    \, \includegraphics[width=\textwidth]{#2}
  \end{minipage} &
  \begin{minipage}[t]{0.5\textwidth}\vspace{0pt}
  \, \, 
  \setlength{\tabcolsep}{6pt}
    \begin{tabular}{@{}ll@{}}
      Name             & #3    \\
      Unique triangles & #5    \\
      Triangles        & #4    \\
      Unique meshes    & #6    \\
      Total meshes     & #7    \\
      Textures         & #8    \\
                       & #9    \\
    \end{tabular}
  \end{minipage} \\
}

\setlength{\tabcolsep}{0.5pt}

\begin{table*}[]
\begin{tabular*}{\textwidth}{lll}
View 1 (Close-up) & View 2 (Overview) & Description \\
\toprule\noalign{\vspace{-0.5\normalbaselineskip}}
\scenerow{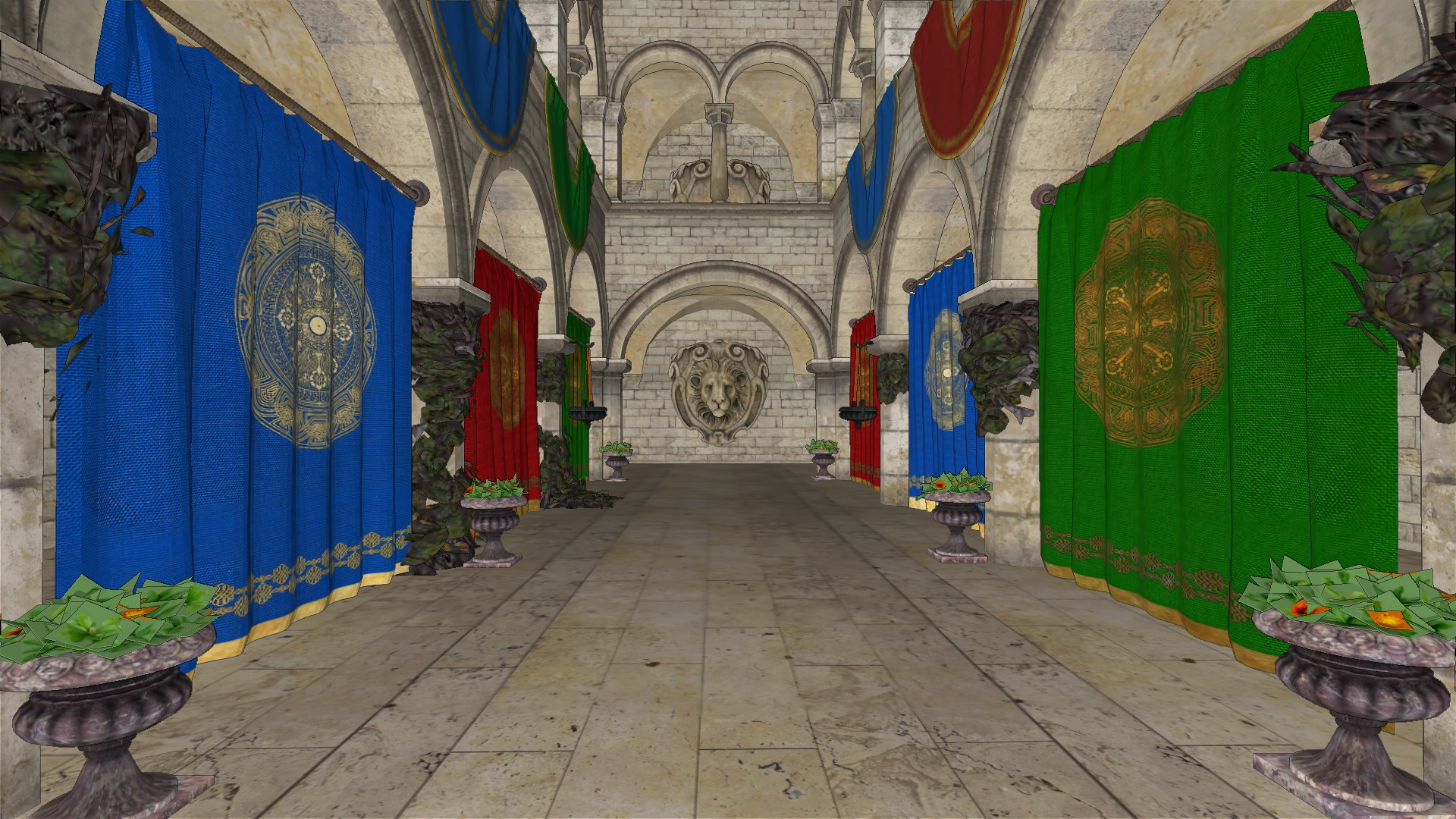}{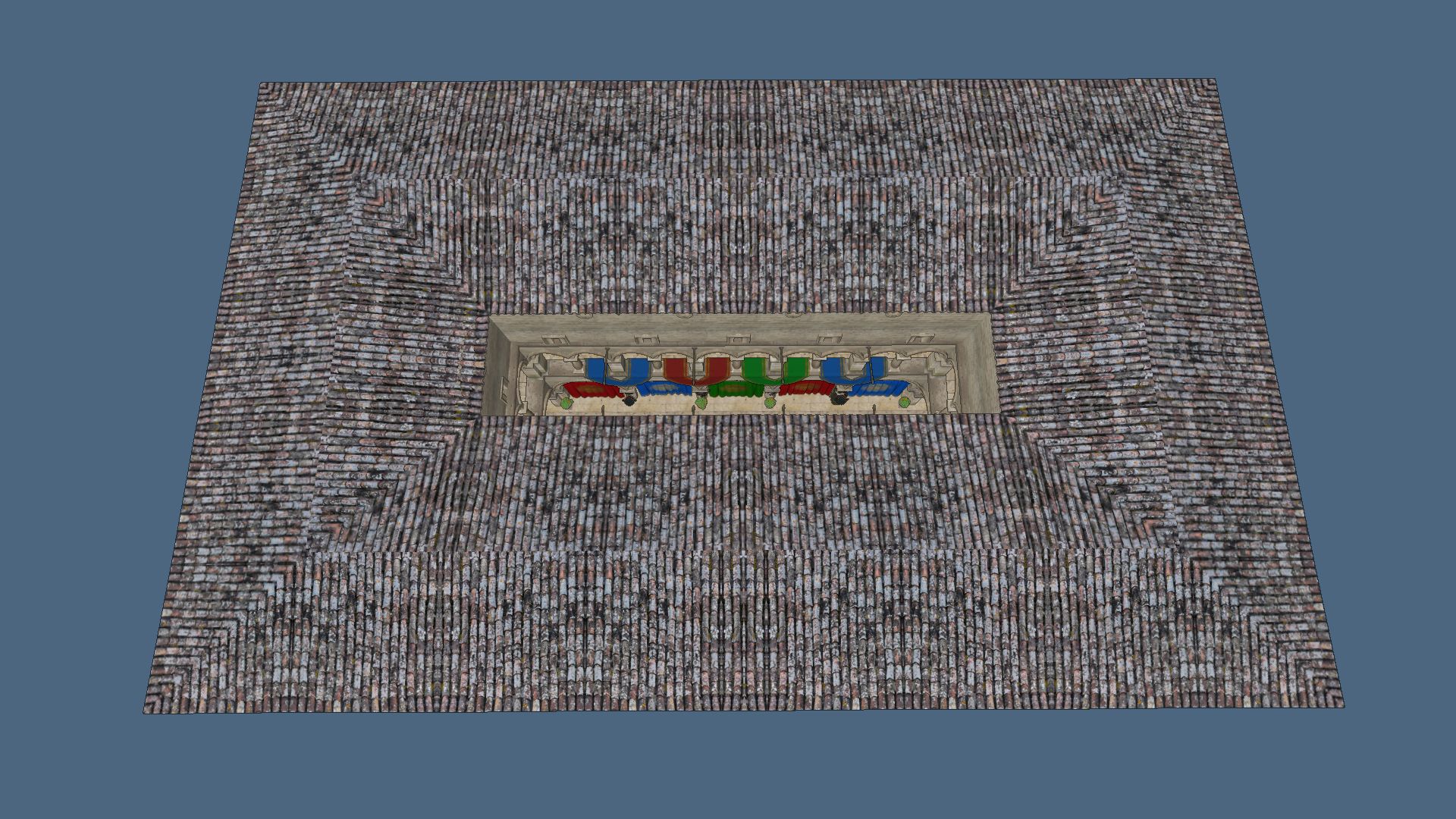}{Sponza}{262k}{262k}{103}{103}{64 x 1k x 1k}{67M Pixels}
\noalign{\vspace{-0.5\normalbaselineskip}}
\scenerow{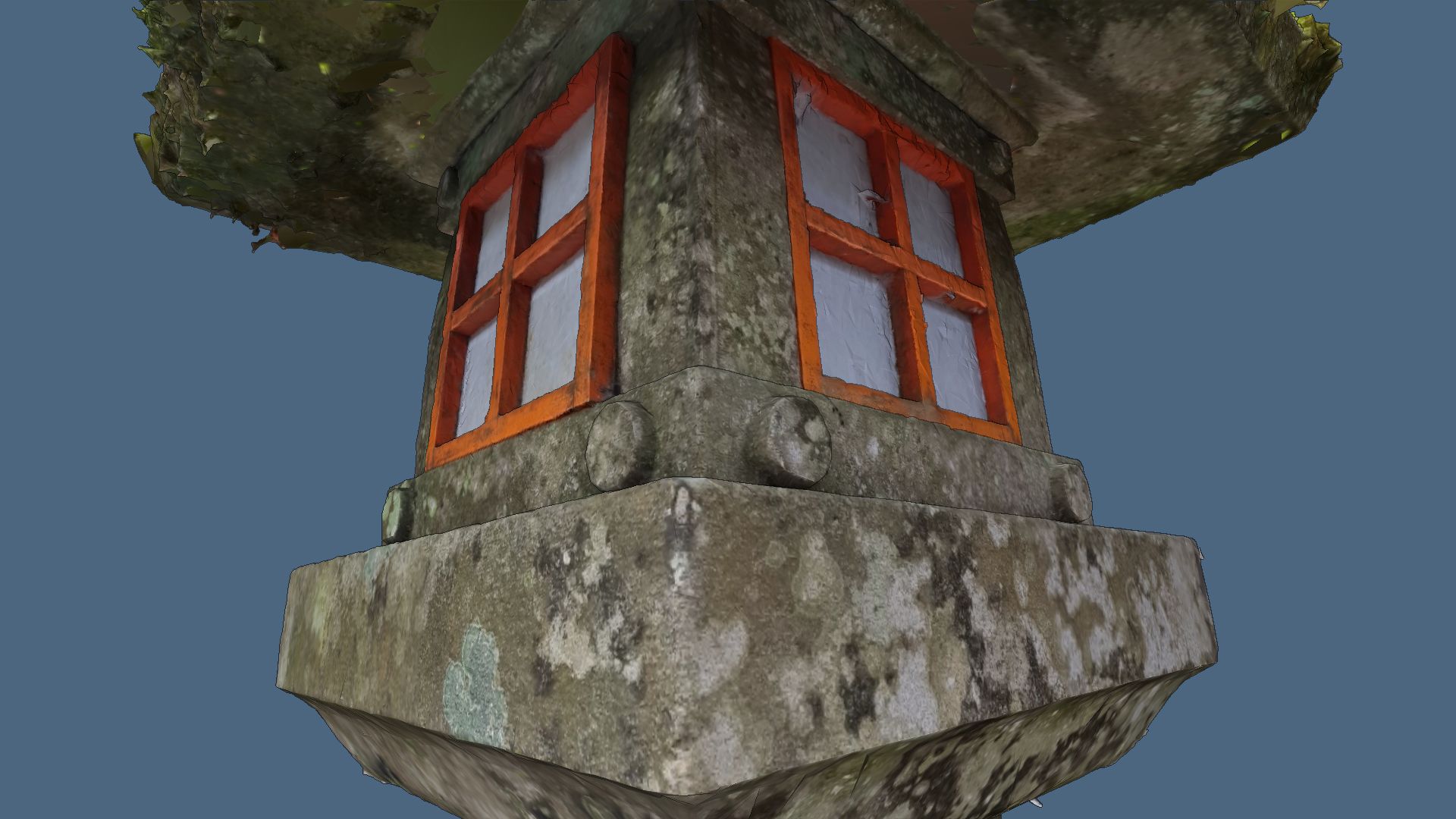}{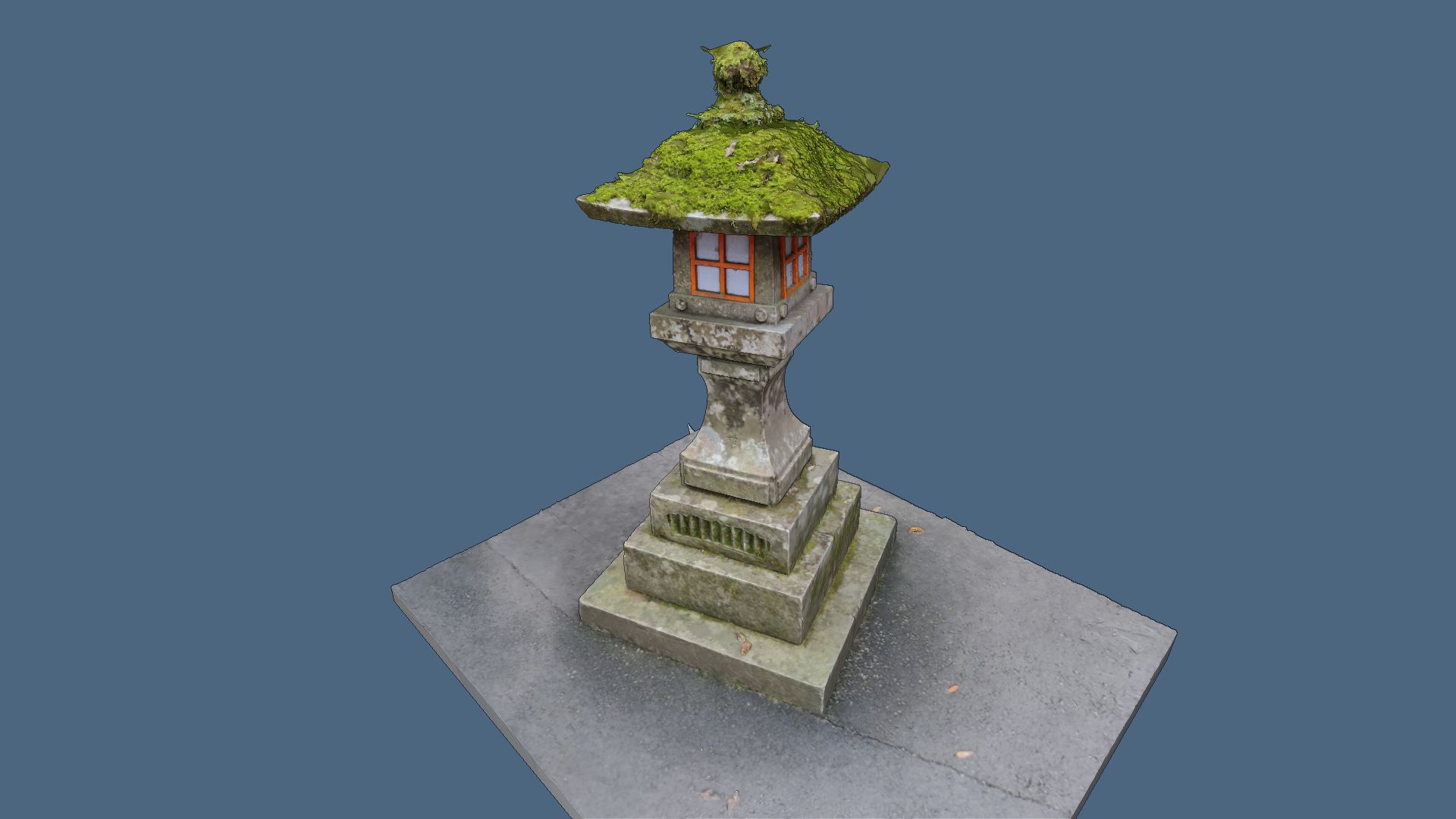}{Lantern}{1 million}{1 million}{1}{1}{1 x 8k}{67M Pixels}
\noalign{\vspace{-0.5\normalbaselineskip}}
\scenerow{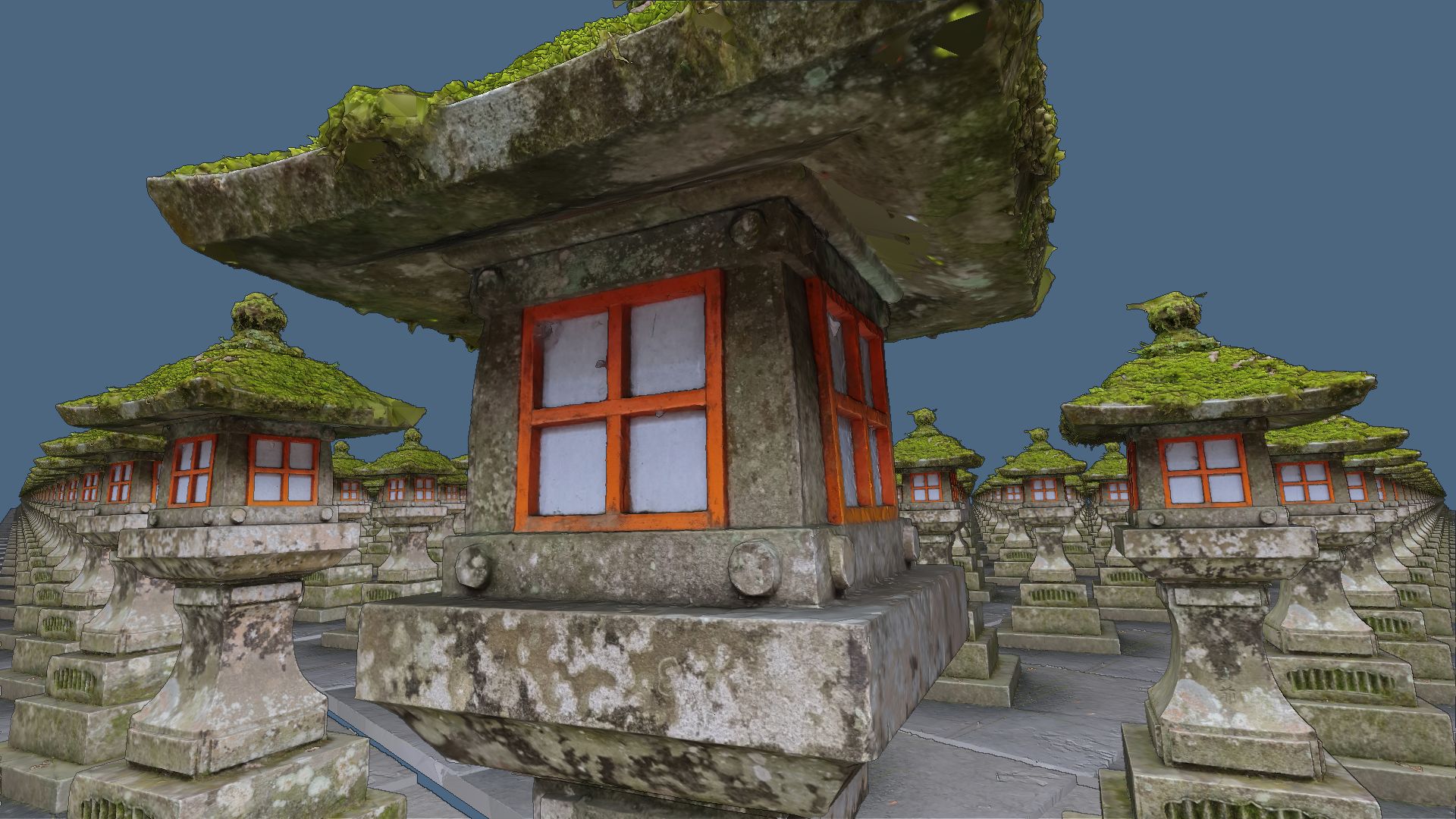}{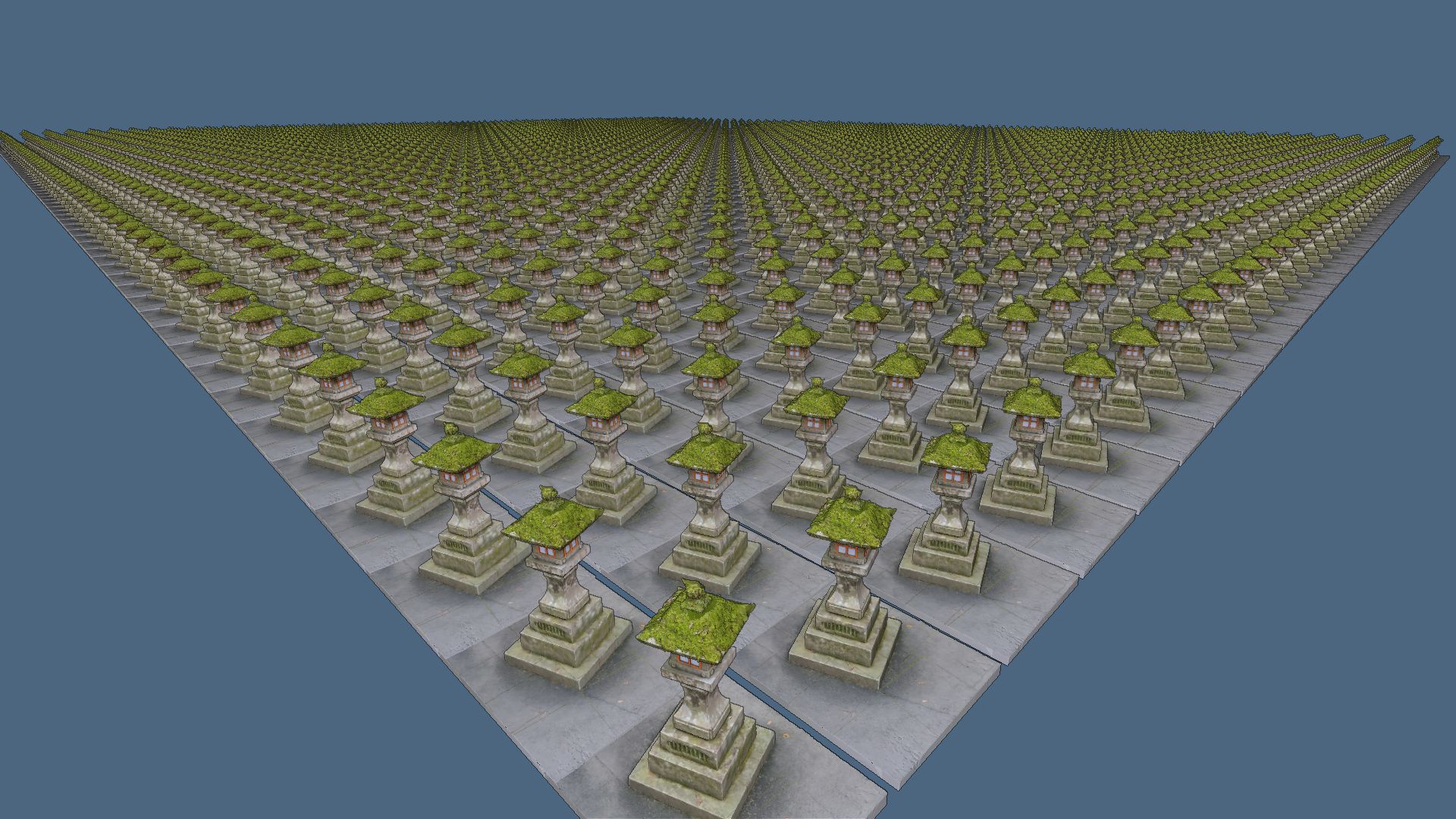}{Lantern Instanced}{3 billion}{1 million}{1}{3000}{1 x 8k}{67M Pixels}
\noalign{\vspace{-0.5\normalbaselineskip}}
\scenerow{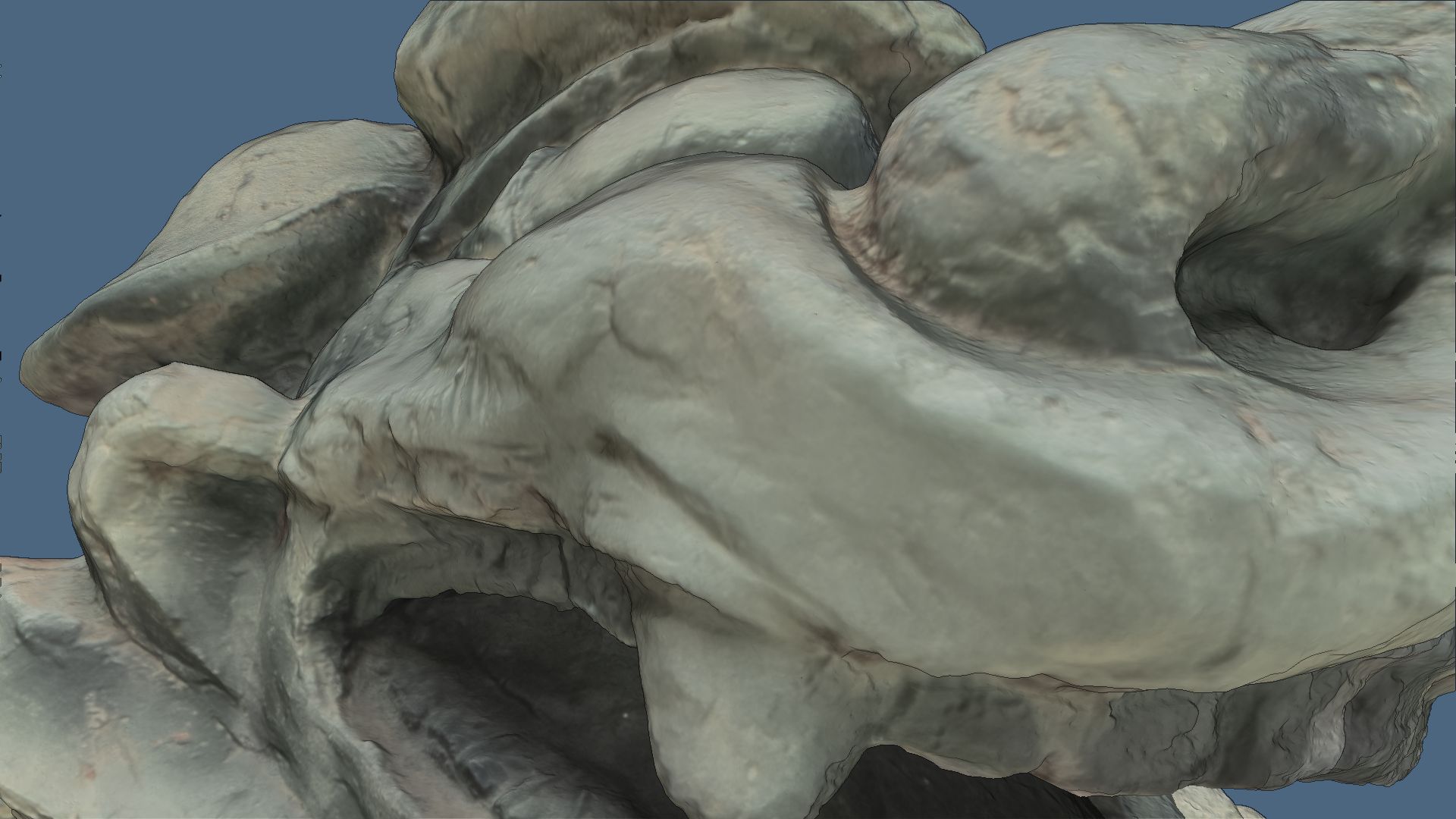}{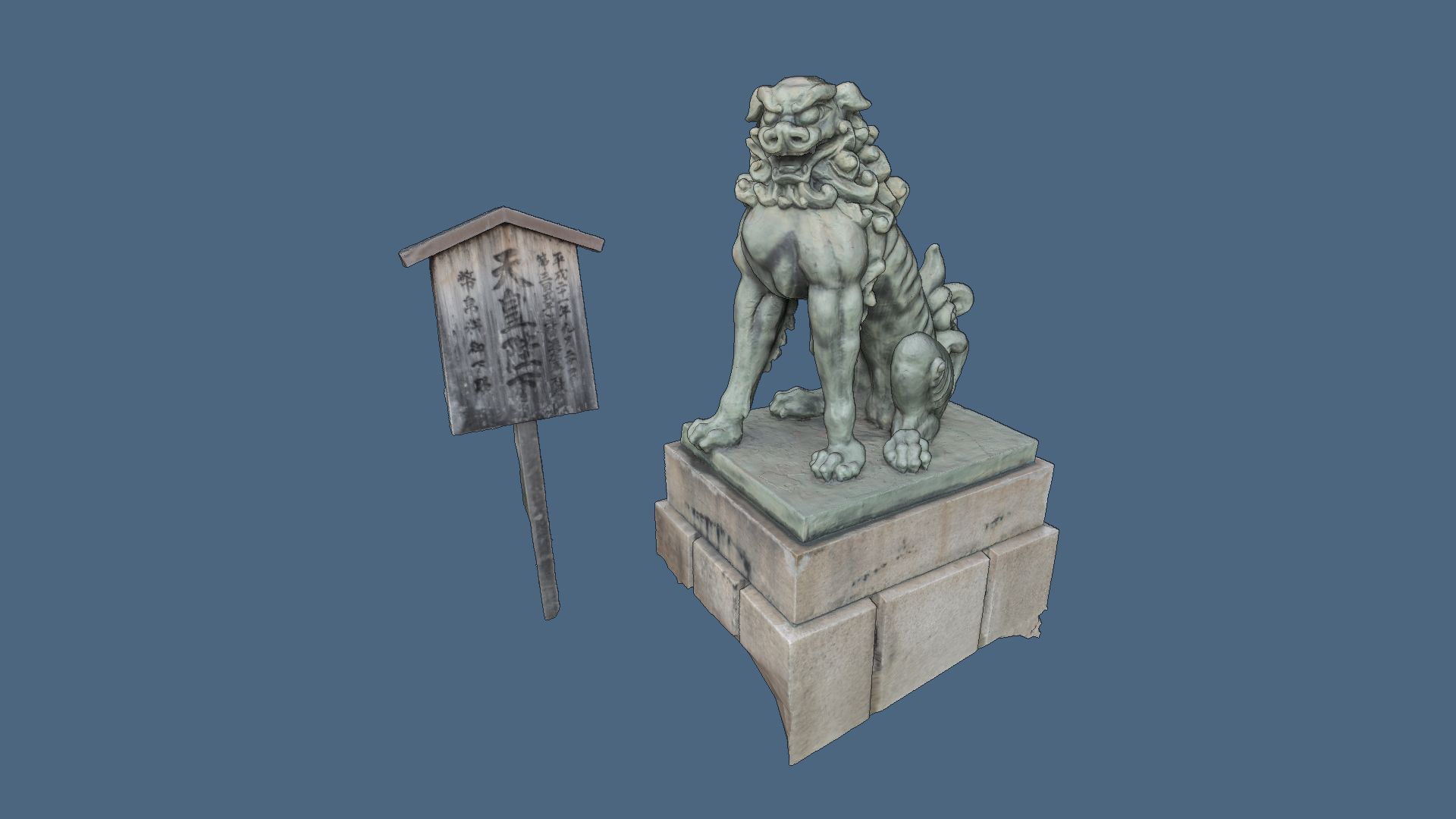}{Komainu Kobe}{60 million}{60 million}{9}{9}{9 x 8k x 8k}{604M Pixels}
\noalign{\vspace{-0.5\normalbaselineskip}}
\scenerow{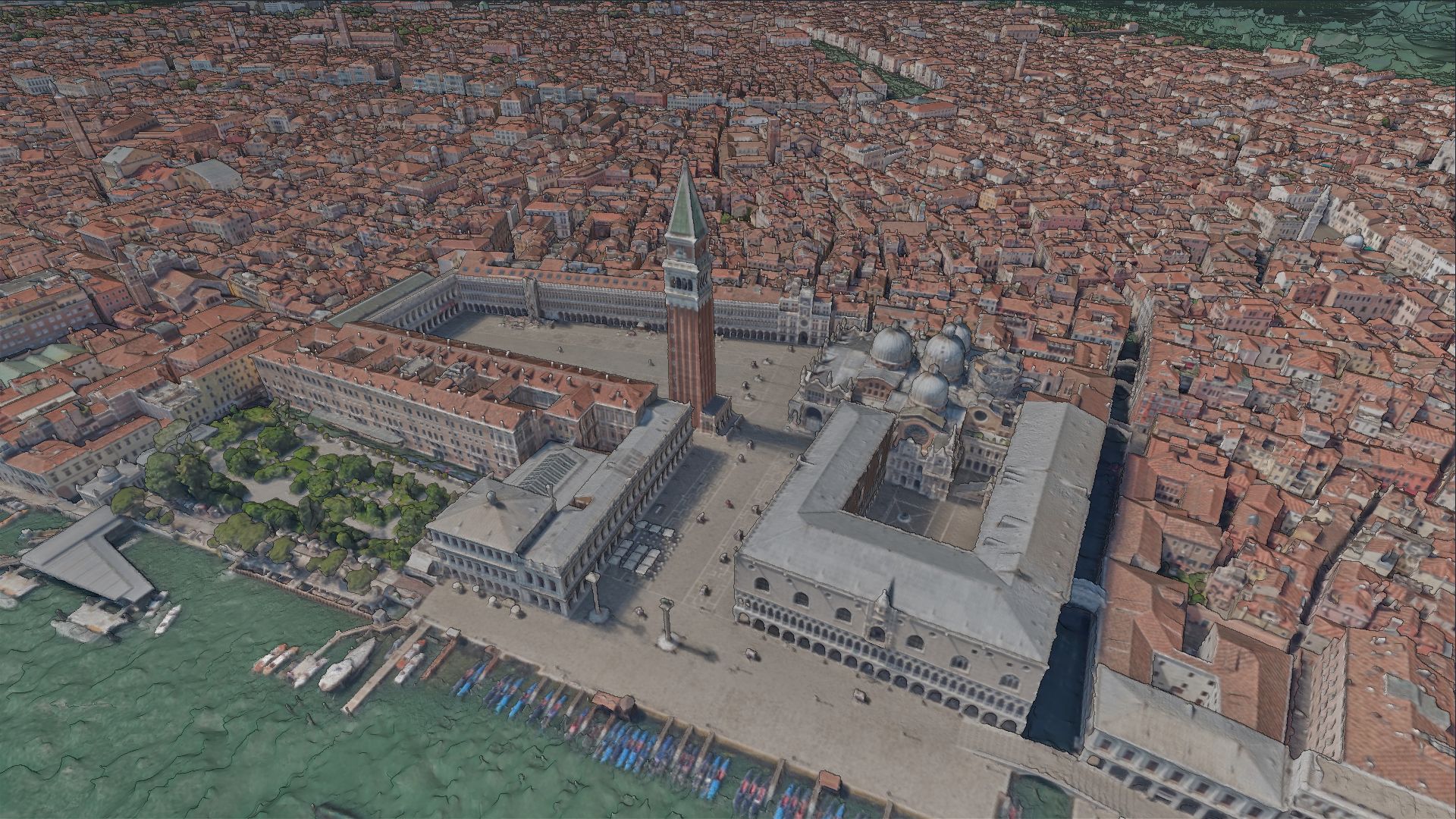}{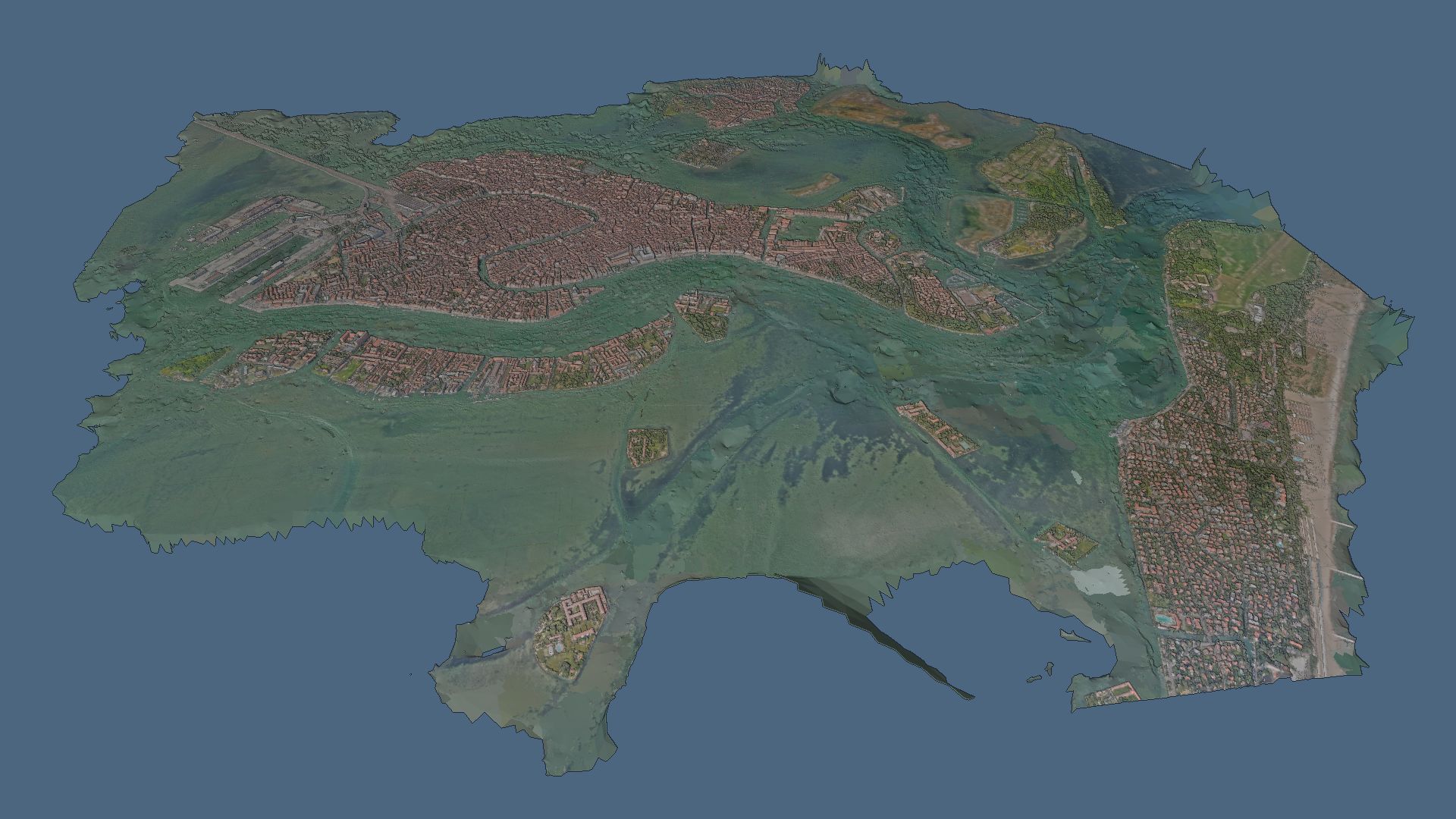}{Venice}{400 million}{400 million}{12}{12}{12 x 16k x 16k}{3.2B Pixels}
\noalign{\vspace{-0.5\normalbaselineskip}}
\scenerow{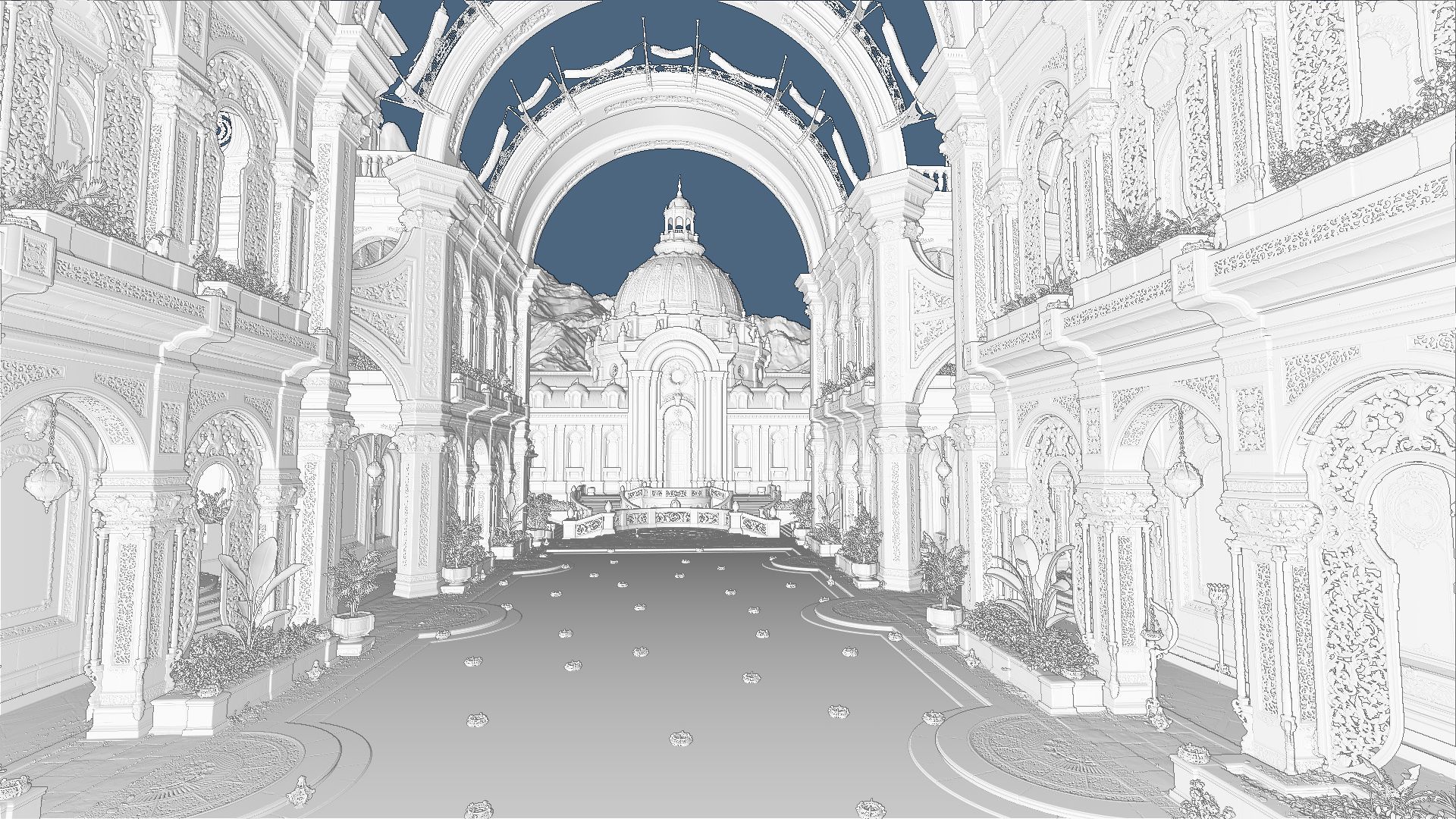}{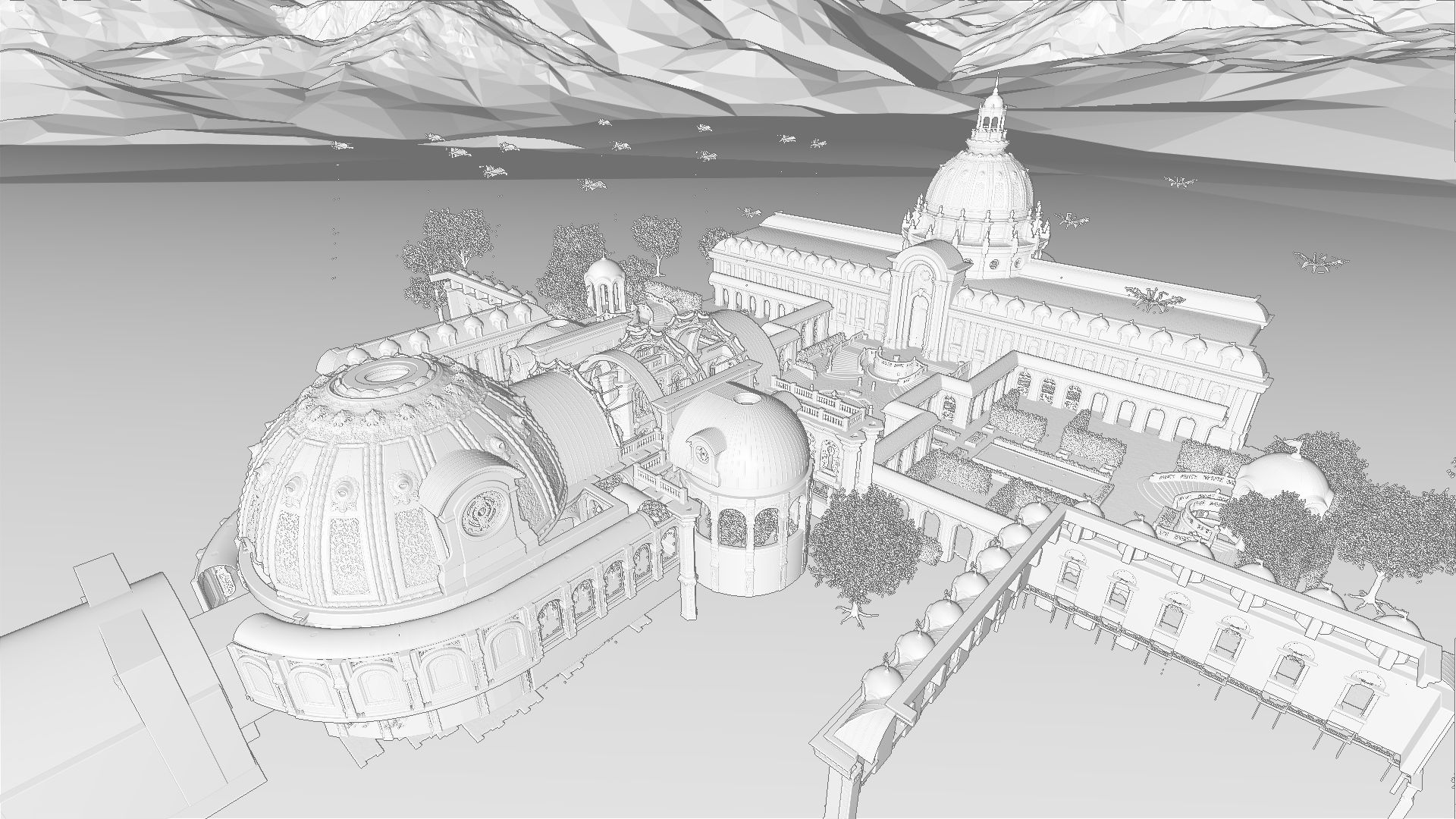}{Zorah}{ 18.9 billion}{1.6 billion}{3880}{26170}{None}{}
\end{tabular*}
\caption{Test data sets. We evaluated scenes ranging from 260k to 18.9 billion triangles, each captured from a close-up and an overview viewpoint. }
\label{tab:test_data}
\end{table*}

The performance of CuRast is compared to a Vulkan-based rendering path, using data sets and configurations as described in Table~\ref{tab:test_data} and Section~\ref{sec:data_and_config}. We focus mainly on large and dense data sets -- the target of our rasterization approach -- but include Sponza as a widely known reference scene.

Benchmarks were captured on following test systems:

\begin{itemize}
    \item \textbf{4070}: CPU: AMD Ryzen 7 5800X; GPU: RTX 4070 12GB
    \item \textbf{4090}: CPU: AMD Ryzen 9 7950X; GPU: RTX 4090 24GB
    \item \textbf{5090}: CPU: AMD Ryzen 9 7950X; GPU: RTX 5090 32GB
\end{itemize}

We compare to two Vulkan-based render paths:

\begin{itemize}
    \item \textbf{VK-ID} (Indexed Draw): Performs one \emph{vkCmdDrawIndexed} per unique mesh, drawing all of its instances. Uses vertex buffers and index buffers.
    \item \textbf{VK-PIP} (Programmable Index Pulling): Performs one \emph{vkCmdDraw} per unique mesh, drawing all of its instances. Instead of vertex and index buffers, this approach uses programmable index pulling and vertex pulling. This approach is needed to draw meshes whose index buffers are compressed to arbitrary bit rates other than 16 or 32 bit. 
\end{itemize}

We omitted a visibility-buffer-based Vulkan pipeline as we observed no difference between these two approaches in our test data sets, except for being slightly slower due to an additional resolve pass. I.e., omitting triangle attributes other than position and indices did not significantly accelerate the geometry rasterization pass in our scenarios. 

During benchmarking of the CUDA and Vulkan approaches, we let each allocate buffers through their own API. Reported timings are the mean over 60 frames.

\subsection{Data Sets and Test Configurations}
\label{sec:data_and_config}

Table~\ref{tab:test_data} depicts the test data sets we use for evaluation. Aside from the amount of geometry, these data sets present a range of challenges that stress different aspects of triangle rasterization pipelines. 

\textbf{Sponza}~\cite{Sponza} comprises 103 low-density meshes that produce large visible triangles. Our rasterizer is not targeted towards this scenario, but it is included due to being a widely used reference scene. 

\textbf{Lantern} is a photogrammetry reconstruction that was simplified to about 1 million triangles. It also serves as a case study for creating thousands of instances (50$\times$60) of a moderately detailed mesh, resulting in a scene with 3 billion triangles.

\textbf{Komainu Kobe}~\cite{JapanStatue} is a photogrammetry reconstruction of one of two lion statues in Kobe / Ikuta-jinja. After a high-resolution reconstruction in Reality Scan, resulting in 284 million triangles, the right Komainu was isolated and simplified to 60 million triangles for use in this evaluation. 

\textbf{Venice}~\cite{Venice} features twelve 16k JPEG textures (3.2 billion pixels, 909MB compressed) that, after decompression and mip map construction, did not fit in GPU memory alongside the geometry. Because the textures are distributed in JPEG format, we decided to keep them in JPEG format for rendering~\cite{JpegTextures}. A lightweight indexing table is constructed so that we are able to randomly access and decode visible JPEG blocks during the resolve pass. Since our Vulkan render path does not support JPEG textures, we also evaluate a scenario with downscaled textures to be able to compare CUDA and Vulkan performances. 

\textbf{Zorah}~\cite{Zorah} is our largest stress-test, featuring a massive 18.9 billion triangles in total, of which 1.6 billion are unique. We use the original v1 release, which has since been replaced by a compressed v2 by NVIDIA. The original release takes up 38.8GB disk space -- 19.7GB for indices, 17.1GB for vertex coordinates, and 2GB for a partial set of uv-coordinates. Since this model has no textures and because most meshes are without uv-coordinates, we ignore the 2GB of uvs. To fit the data set on an RTX 4090, we apply the compression scheme as outlined in Section~\ref{sec:compression}. After compression, index and vertex buffers require 13.1GB and 8.6GB of memory. We filter out a couple of billboards for aesthetic reasons, particularly those labeled "FogCard" and "Plane". 

To ensure a fair comparison between our software rasterizer and Vulkan indexed draw calls, we also apply meshoptimizer/gltfpack~\cite{meshoptimizer} to some scenarios. Meshoptimizer rearranges vertices and triangles to improve locality and helps GPUs to reuse vertices. 

Since indexed draw calls in Vulkan only support 16 or 32 bit precision index buffers, test scenarios with compression are only evaluated in CUDA and VK-PIP, but not in VK-ID. VK-PIP supports compressed indices by loading and decoding them inside the vertex shader. 

CUDA timings are measured with CUDA events, and Vulkan timings are measured via timestamp queries. CUDA timings in Table~\ref{tab:benchmark_results} cover stages 1, 2, and 3, and the resolve pass. Vulkan timings cover the vkCmdDraw and vkCmdDrawIndexed commands. 

\newcommand{\fa}[1]{{\textcolor[HTML]{1a9850}{#1}}} 
\newcommand{\sl}[1]{{\textcolor[HTML]{d73027}{#1}}} 

\newcolumntype{Y}{>{\raggedleft\arraybackslash}X}

\begin{table*}[t]
\centering
\caption{Benchmark timings in milliseconds. \textbf{c}: Compressed coordinates and index buffers. \textbf{m}: Meshes optimized with meshoptimizer/gltfpack~\cite{meshoptimizer}. \textbf{j}: JPEG-compressed textures. \textbf{h}: Resized texture to half resolution. \textbf{--}: Out of memory. \textbf{$\times$}: Unsupported test configuration. E.g., Vulkan indexed draws do not support compressed index buffers, and JPEG-compressed textures are only supported in our CUDA rasterizer.}
\label{tab:benchmark_results}
\rowcolors{1}{white}{gray!10}
\begin{tabularx}{\textwidth}{
        l<{\hspace{5pt}}
        l<{\hspace{10pt}}
        r<{\hspace{10pt}}
        l l l l
        *{9}{Y}}
        \toprule
        & & & & & &
        & \multicolumn{3}{c}{RTX 4070}
        & \multicolumn{3}{c}{RTX 4090}
        & \multicolumn{3}{c}{RTX 5090} \\
        \cmidrule(lr){8-10} \cmidrule(lr){11-13} \cmidrule(lr){14-16}
        & Scene & vis. tris & {} & {} & {} &
        & {CuRast} & {VK-ID} & {VK-PIP}
        & {CuRast} & {VK-ID} & {VK-PIP}
        & {CuRast} & {VK-ID} & {VK-PIP} \\
        \midrule

        %
        %
        & Sponza               &  197.0k &   &   &   &   &    \sl{0.580} &    \fa{0.068} &         0.071 &    \sl{0.271} &    \fa{0.040} &         0.052 &    \sl{0.251} &    \fa{0.022} &         0.025 \\
        & Lantern              &    1.0M &   & m &   &   &    \sl{0.460} &    \fa{0.121} &         0.140 &    \sl{0.208} &    \fa{0.056} &         0.094 &    \sl{0.162} &    \fa{0.042} &         0.049 \\
        & Lantern Instances    &    3.1B &   & m &   &   &   \fa{41.351} &       241.833 &  \sl{367.267} &   \fa{17.325} &       142.001 &  \sl{282.385} &    \fa{9.951} &       125.677 &  \sl{142.147} \\
        & Komainu Kobe         &   28.7M &   &   &   &   &    \fa{1.855} &    \sl{3.975} &         3.627 &    \fa{0.865} &    \sl{2.791} &         2.678 &    \fa{0.644} &    \sl{2.413} &         1.365 \\
        & Komainu Kobe         &   28.7M &   & m &   &   &         1.656 &    \fa{1.524} &    \sl{3.466} &    \fa{0.825} &         1.380 &    \sl{2.640} &    \fa{0.544} &         1.220 &    \sl{1.323} \\
        & Venice               &  399.9M & c & m &   & h &            -- &            -- &            -- &    \fa{6.929} &      $\times$ &   \sl{37.571} &    \fa{4.681} &      $\times$ &   \sl{18.892} \\
        & Venice               &  399.9M & c &   &   & h &            -- &            -- &            -- &   \fa{10.126} &      $\times$ &   \sl{37.763} &    \fa{7.288} &      $\times$ &   \sl{19.048} \\
        & Venice               &  399.9M &   &   & j &   &            -- &            -- &            -- &   \fa{10.732} &      $\times$ &      $\times$ &    \fa{7.692} &      $\times$ &      $\times$ \\
        & Venice               &  399.9M &   & m & j &   &            -- &            -- &            -- &    \fa{9.548} &      $\times$ &      $\times$ &    \fa{5.242} &      $\times$ &      $\times$ \\
        & Venice               &  399.9M &   & m &   & h &            -- &            -- &            -- &    \fa{9.570} &        19.710 &   \sl{37.575} &    \fa{5.252} &        17.106 &   \sl{18.888} \\
        & Venice               &  399.9M &   &   &   & h &            -- &            -- &            -- &   \fa{10.724} &   \sl{49.616} &        37.668 &    \fa{8.051} &   \sl{42.387} &        19.333 \\
        & Zorah                &   13.6B & c &   &   &   &            -- &            -- &            -- &   \fa{74.906} &      $\times$ & \sl{1778.303} &   \fa{57.569} &      $\times$ &  \sl{633.081} \\
        \multirow{-13}{*}{\rotatebox{90}{closeup}}
        & Zorah                &   13.6B & c & m &   &   &            -- &            -- &            -- &   \fa{69.764} &      $\times$ & \sl{1250.210} &   \fa{53.395} &      $\times$ &  \sl{631.095} \\

        \midrule

        %
        %
        & Sponza               &  262.3k &   &   &   &   &    \sl{0.439} &    \fa{0.042} &         0.065 &    \sl{0.224} &    \fa{0.029} &         0.052 &    \sl{0.239} &    \fa{0.018} &         0.022 \\
        & Lantern              &    1.0M &   & m &   &   &    \sl{0.358} &    \fa{0.113} &         0.130 &    \sl{0.157} &    \fa{0.056} &         0.094 &    \sl{0.128} &    \fa{0.042} &         0.050 \\
        & Lantern Instances    &    3.1B &   & m &   &   &   \fa{36.550} &       245.304 &  \sl{367.131} &   \fa{15.516} &       141.747 &  \sl{282.331} &    \fa{9.614} &       125.680 &  \sl{142.131} \\
        & Komainu Kobe         &   60.0M &   &   &   &   &    \fa{3.376} &    \sl{8.338} &         7.482 &    \fa{1.568} &    \sl{6.100} &         5.870 &    \fa{1.134} &    \sl{5.026} &         2.800 \\
        & Komainu Kobe         &   60.0M &   & m &   &   &    \fa{2.963} &         5.158 &    \sl{7.293} &    \fa{1.456} &         2.881 &    \sl{5.863} &    \fa{0.932} &         2.560 &    \sl{2.766} \\
        & Venice               &  399.9M & c & m &   & h &            -- &            -- &            -- &    \fa{7.271} &      $\times$ &   \sl{37.571} &    \fa{4.985} &      $\times$ &   \sl{18.890} \\
        & Venice               &  399.9M & c &   &   & h &            -- &            -- &            -- &    \fa{9.955} &      $\times$ &   \sl{37.516} &    \fa{7.258} &      $\times$ &   \sl{19.018} \\
        & Venice               &  399.9M &   &   & j &   &            -- &            -- &            -- &   \fa{10.948} &      $\times$ &      $\times$ &    \fa{7.757} &      $\times$ &      $\times$ \\
        & Venice               &  399.9M &   & m & j &   &            -- &            -- &            -- &    \fa{9.759} &      $\times$ &      $\times$ &    \fa{5.377} &      $\times$ &      $\times$ \\
        & Venice               &  399.9M &   & m &   & h &            -- &            -- &            -- &    \fa{9.766} &        19.690 &   \sl{37.580} &    \fa{5.380} &        17.103 &   \sl{18.882} \\
        & Venice               &  399.9M &   &   &   & h &            -- &            -- &            -- &   \fa{10.946} &   \sl{49.596} &        37.595 &    \fa{7.789} &   \sl{42.344} &        19.097 \\
        & Zorah                &   18.8B & c &   &   &   &            -- &            -- &            -- &   \fa{99.698} &      $\times$ & \sl{2388.751} &   \fa{76.532} &      $\times$ &  \sl{873.428} \\
        \multirow{-13}{*}{\rotatebox{90}{overview}}
        & Zorah                &   18.8B & c & m &   &   &            -- &            -- &            -- &   \fa{93.043} &      $\times$ & \sl{1729.774} &   \fa{70.667} &      $\times$ &  \sl{872.477} \\

        \bottomrule
\end{tabularx}
\end{table*}

\begin{figure}[htp]
    \centering
    \includegraphics[width=\columnwidth]{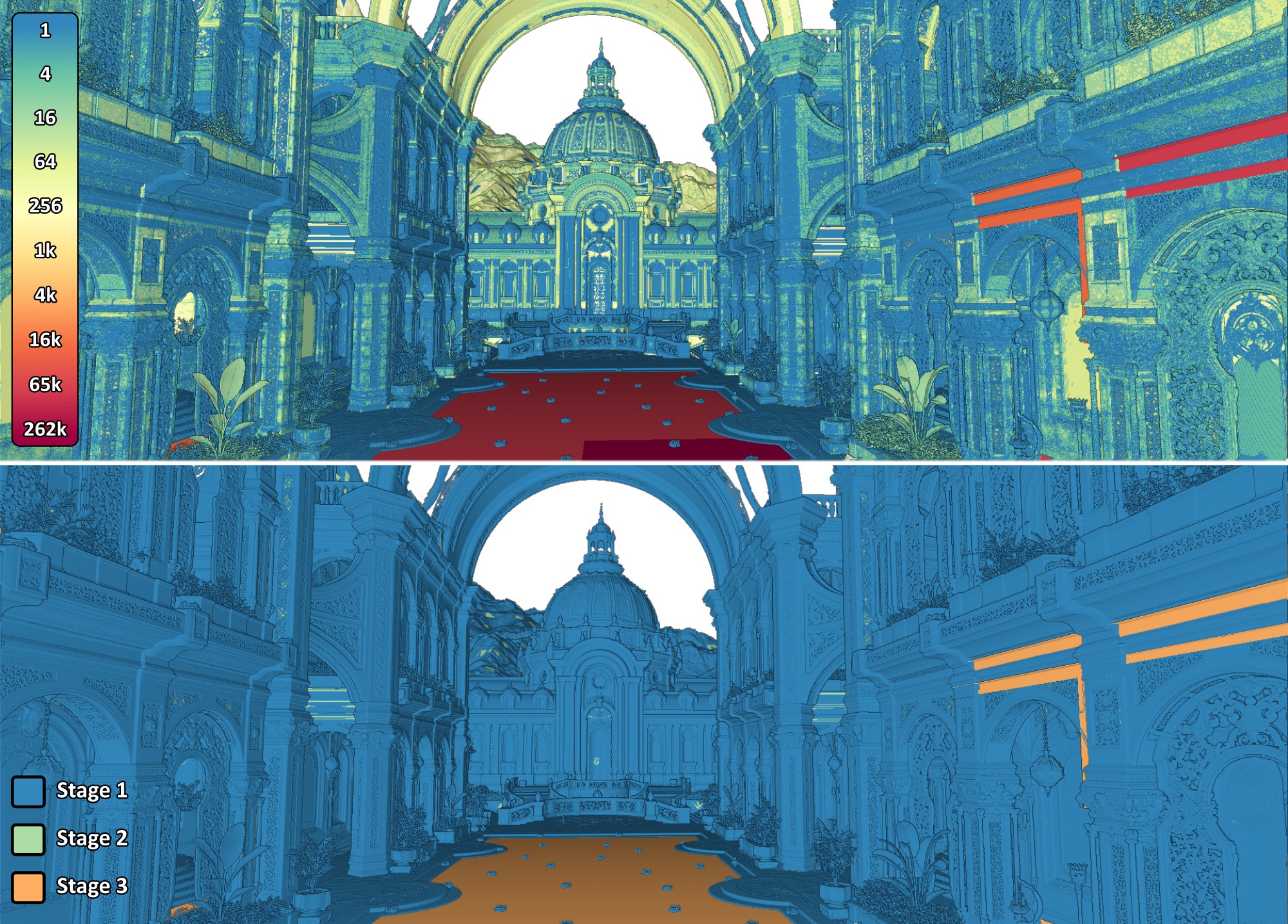}
    \caption{Top: Colored by size (px) of the bounding box of a pixel's triangle. Bottom: Pixels colored by the rasterization stage that created the fragment. In Zorah, the vast majority of triangles are rasterized in stage 1.}
    \label{fig:zorah_stages}
\end{figure}

Table~\ref{tab:benchmark_results} shows the benchmarking results of various scenarios on three GPUs. Figure~\ref{fig:zorah_stages} illustrates the density of triangles and the stages used to rasterize them. In Sponza, Vulkan indexed draws are 7 to 13 times faster than our approach. The Lantern scene shortens that gap to a factor of about 4. For the remaining scenes, CuRast performs significantly faster than Vulkan. The biggest increases are seen in scenes with numerous instances (Lantern Instances, Zorah), where CuRast only loads triangle geometry once to rasterize all instances of these triangles. CUDA also performs multiple times faster than Vulkan indexed draws in scenarios without meshoptimized geometry, although that gap to indexed draws shortens to a factor of $2\times$ after meshoptimization. 

Summary of notable observations based on the benchmark results:

\begin{itemize}
    \item CuRast thrives with large models and numerous instances of large models. 
    \item Meshoptimization has the biggest impact on Vulkan instanced draw calls, often doubling the performance when rendering dense meshes. The impact on CuRast or Vulkan programmable index pulling is significant but comparatively modest. 
    \item The RTX 5090 shows major improvements for programmable index pulling, making it nearly twice as fast in all cases compared to the RTX 4090. Improvements for indexed draws have been smaller. This makes index pulling significantly faster than indexed draw in some non-meshoptimized scenarios. After mesh optimization, indexed draws are always faster than index pulling, but the difference is now small. 
\end{itemize}

\subsubsection{Performance of Individual Stages}

Table~\ref{tab:timings_stages} breaks down the timings for the individual stages, measured on an RTX 4090.

\begin{table}[h]
    \centering
    \caption{Timings (ms) of individual rasterization stages and resolve on an RTX 4090. }
    \label{tab:timings_stages}
    \begin{tabular*}{\columnwidth}{@{\extracolsep{\fill}} l l r r r r}
        \hline
        \textbf{} & & \textbf{Stage 1} & \textbf{Stage 2} & \textbf{Stage 3} & \textbf{Resolve} \\
        \hline
        Sponza (closeup)   &   &  0.048 & 0.054 & 0.063 & 0.098 \\
        Venice (closeup)   & h & 10.162 & 0.033 & 0.006 & 0.517 \\
        Venice (overview)  & h & 10.186 & 0.013 & 0.007 & 0.734 \\
        Zorah (closeup)    & c & 74.109 & 0.193 & 0.183 & 0.416 \\
        Zorah (overview)   & c & 99.067 & 0.130 & 0.210 & 0.254 \\
        \hline
    \end{tabular*}
\end{table}

\subsubsection{Culling tiny triangles}

In massively dense scenes, a large number of triangles end up in between pixel samples without intersecting them. Culling these early affects the performance of the stage 1 kernel as shown in Table~\ref{tab:culling_tiny}. Our takeaway here is that it substantially benefits the most massive of data sets, without negatively affecting small data sets that do not benefit from this optimization.

\begin{table}[h]
    \centering
    \caption{Early-culling triangles that do not intersect a sample.}
    \label{tab:culling_tiny}
    \begin{tabular*}{\columnwidth}{@{\extracolsep{\fill}} l c r r}
        \hline
        \textbf{} & & \textbf{enabled} & \textbf{disabled} \\
        \hline
        Sponza (closeup)  &   & 0.048 ms  &  0.048 ms \\
        Venice (overview) & h & 10.1 ms  &  10.2 ms \\
        Zorah (closeup)   & c & 74.1 ms & 102.5 ms \\
        Zorah (overview)  & c & 99.1 ms & 140.2 ms \\
        \hline
    \end{tabular*}
\end{table}

\subsection{Anti-Aliasing (Super Sampling)}

We support anti-aliasing via super sampling, which affects the performance as shown in Table~\ref{tab:supersampling}. For Sponza, rendering duration increases nearly equal to sample count, while for Venice and Zorah the rendering time increases at a much lower rate. 

\begin{table}[H]
    \centering
    \caption{Super Sampling performance (ms) on an RTX 5090.}
    \label{tab:supersampling}
    \begin{tabular*}{\columnwidth}{@{\extracolsep{\fill}} l l r r r}
        \hline
        \textbf{} & & \textbf{No SS} & \textbf{2x2 SS} & \textbf{4x4 SS} \\
        \hline
        Sponza (closeup)  &   &  0.251 &  0.847 &  2.968 \\
        Venice (overview) & h &  7.789 &  8.845 & 12.261 \\
        Zorah (closeup)   & c & 57.569 & 64.913 & 84.008 \\
        \hline
    \end{tabular*}
\end{table}

\section{Conclusion and Future Work}

In this paper, we have shown that a software rasterization pipeline specifically targeted towards dense, opaque triangle meshes can outperform the native hardware rasterization pipeline by factors of 2-12. CuRast mainly benefits from scenarios such as fewer meshes with large triangle counts producing pixel-sized triangles; instancing of dense meshes; and, to a smaller extent, from geometry compression. Vulkan, on the other hand, remains the champion when it comes to dealing with numerous small meshes. 

Based on the results, we believe that software rasterization is at a stage where it is already a viable option for use cases such as reconstruction of massive photogrammetry data sets with users in the loop for manual editing, or generally editing large, dense meshes. 

For more general graphics applications, and games in particular, future work will need to improve performance for large numbers of small meshes and add support for features like blending and transparency. Given the success of 3DGS, the latter should be easily doable in CUDA. We also believe that past and ongoing research on clustered approaches and/or hierarchical LODs~\cite{karis2021nanite,RTXMegaGeometry,meshoptimizer} is the way forward for games, and integrating these approaches into CuRast is likely to result in major improvements in rendering performance. Nanite, taking advantage of clustered geometry, also implements an interesting mechanism akin to vertex-reuse. Instead of simply launching one thread per triangle and independently processing potentially shared vertices multiple times as we do, they first process vertices in a cluster using one thread per vertex, store the results in shared memory, and then process the triangles using one thread per triangle~\cite{NaniteReyes}. Kuth et al.~\cite{KUTH2025104292} propose a meshlet compression approach that can reduce index buffer sizes down to 9 bit per triangle, thus reducing pressure on memory bandwidth. Finally, limiting processing to fragments within the triangle instead of the triangle's screen-space bounding box is a likely way to improve performance. While we did not yet have success in doing so, this may be a matter of crafting the right implementation. It may also be more advantageous for less dense geometry than the data sets we were targeting in this work. 

\section{Acknowledgments}


The authors wish to thank following data set providers: 
Keenan Crane for the \emph{Spot} model~\cite{crane2013robust}; %
Marko Dabrovic, Frank Meinl, Hans-Kristian Arntzen, Morgan McGuire, Crytek and Ludicon for \emph{Sponza} in glb format with lossless-compressed textures~\cite{Sponza}; %
Gildas Sidobre and the NRHK for the images of the Komainu Kobe~\cite{JapanStatue}; %
Iconem and the Fondazione Musei Civici di Venezia for the Venice data set~\cite{Venice}; %
NVIDIA for the \emph{Zorah} data set~\cite{Zorah}.

This research has been funded by WWTF project \emph{ICT22-055 - Instant Visualization and Interaction for Large Point Clouds} and WWTF project \emph{ICT25-084 - Instant Visualization and Editing of Arbitrarily Large 3D Data Sets}. 

\printbibliography

@String{tog = "ACM TOG"}

@misc{karis2021nanite,
    title        = {A Deep Dive into Nanite Virtualized Geometry},
    author       = {Karis, Brian and Stubbe, Rune and Wihlidal, Graham},
    year         = {2021},
    howpublished = {SIGGRAPH 2021 Course: Advances in Real-Time Rendering in Games},
    note         = {Industry talk},
    url          = {https://advances.realtimerendering.com/s2021/Karis_Nanite_SIGGRAPH_Advances_2021_final.pdf}
}

@article{VisibilityBuffer,
    author  = {Christopher A. Burns and Warren A. Hunt},
    title   = {The Visibility Buffer: A Cache-Friendly Approach to Deferred Shading},
    year    = {2013},
    month   = {August},
    day     = {12},
    journal = {Journal of Computer Graphics Techniques (JCGT)},
    volume  = {2},
    number  = {2},
    pages   = {55--69},
    url     = {http://jcgt.org/published/0002/02/04/},
    issn    = {2331-7418}
}

@inproceedings{VisibilityBuffer2,
    title={Deferred attribute interpolation for memory-efficient deferred shading},
    author={Schied, Christoph and Dachsbacher, Carsten},
    booktitle={Proceedings of the 7th Conference on High-Performance Graphics},
    pages={43--49},
    year={2015}
}

@misc{Zorah,
    title        = {Zorah},
    author       = {NVIDIA},
    year         = {2025},
    howpublished = {NVIDIA nvpro-samples},
    note         = {Export of NVIDIA RTX Kit - Zorah Sample as presented in "NVIDIA RTX Advances with Neural Rendering and Digital Human Technologies at GDC 2025"},
    url          = {https://github.com/nvpro-samples/vk_lod_clusters/blob/main/README.md#zorah-demo-scene}
}

@misc{Venice,
    title        = {Venice},
    author       = {Iconem and the Fondazione Musei Civici di Venezia},
    year         = {2026},
    url          = {https://iconem.com/}
}

@misc{ZorahGDC2025,
    title        = {Visualizing Next-Gen Games With RTX Neural Rendering and Unreal Engine 5},
    author       = {NVIDIA},
    year         = {2025},
    howpublished = {GDC 2025},
    url          = {https://www.youtube.com/watch?v=udqApkIqZmQ}
}

@misc{meshoptimizer,
  author       = {Kapoulkine, Arseny},
  title        = {Meshoptimizer / gltfpack v1.1},
  year         = {2026},
  month        = apr,
  day          = {2},
  howpublished = {\url{https://github.com/zeux/meshoptimizer}},
  note         = {}
}

@inproceedings{liu2010freepipe,
    author = {Liu, Fang and Huang, Meng-Cheng and Liu, Xue-Hui and Wu, En-Hua},
    title = {FreePipe: a programmable parallel rendering architecture for efficient multi-fragment effects},
    year = {2010},
    isbn = {9781605589398},
    publisher = {Association for Computing Machinery},
    address = {New York, NY, USA},
    url = {https://doi.org/10.1145/1730804.1730817},
    doi = {10.1145/1730804.1730817},
    booktitle = {Proceedings of the 2010 ACM SIGGRAPH Symposium on Interactive 3D Graphics and Games},
    pages = {75–82},
    numpages = {8},
    location = {Washington, D.C.},
    series = {I3D '10}
}

@book{catmull1974subdivision,
  title={A subdivision algorithm for computer display of curved surfaces},
  author={Catmull, Edwin Earl},
  year={1974},
  publisher={The University of Utah}
}

@inproceedings{10.1145/2018323.2018337,
author = {Laine, Samuli and Karras, Tero},
title = {High-performance software rasterization on GPUs},
year = {2011},
isbn = {9781450308960},
publisher = {Association for Computing Machinery},
address = {New York, NY, USA},
url = {https://doi.org/10.1145/2018323.2018337},
doi = {10.1145/2018323.2018337},
booktitle = {Proceedings of the ACM SIGGRAPH Symposium on High Performance Graphics},
pages = {79–88},
numpages = {10},
location = {Vancouver, British Columbia, Canada},
series = {HPG '11}
}

@article{Kenzel:2018:CURE,
author = {Kenzel, Michael and Kerbl, Bernhard and Schmalstieg, Dieter and Steinberger, Markus},
title = {A High-Performance Software Graphics Pipeline Architecture for the GPU},
journal = {ACM Trans. Graph.},
issue_date = {November 2018},
volume = {37}, number = {4},
month = nov,
year = {2018},
articleno = {140},
numpages = {15},
doi = {10.1145/3197517.3201374},
publisher = {ACM},
address = {New York, NY, USA},
}

@article{KUTH2025104292,
    title = {Real-time meshlet decompression},
    journal = {Computers \& Graphics},
    volume = {131},
    pages = {104292},
    year = {2025},
    issn = {0097-8493},
    doi = {https://doi.org/10.1016/j.cag.2025.104292},
    url = {https://www.sciencedirect.com/science/article/pii/S0097849325001335},
    author = {Bastian Kuth and Max Oberberger and Felix Kawala and Sander Reitter and Sebastian Michel and Matthäus Chajdas and Quirin Meyer},
}

@inproceedings{10.1145/54852.378457,
author = {Pineda, Juan},
title = {A parallel algorithm for polygon rasterization},
year = {1988},
isbn = {0897912756},
publisher = {Association for Computing Machinery},
address = {New York, NY, USA},
url = {https://doi.org/10.1145/54852.378457},
doi = {10.1145/54852.378457},
booktitle = {Proceedings of the 15th Annual Conference on Computer Graphics and Interactive Techniques},
pages = {17–20},
numpages = {4},
series = {SIGGRAPH '88}
}

@article{Marrs2018Shadows,
    author =       {Adam Marrs and Benjamin Watson and Christopher Healey}, 
    title =        {View-warped Multi-view Soft Shadows for Local Area Lights},
    year =         2018,
    day =          26,
    journal =      {Journal of Computer Graphics Techniques (JCGT)},
    volume =       7,
    number =       3,
    pages =        {1--28},
}

@incollection{2015learning,
   author    = {Evans, Alex},
   title     = {{Learning from failure: A Survey of Promising, Unconventional and Mostly Abandoned Renderers for ‘Dreams PS4’, a Geometrically Dense, Painterly UGC Game}},
   booktitle = {ACM SIGGRAPH 2015 Courses, Advances in Real-Time Rendering in Games},
   year      = {2015},
    note = {\url{https://advances.realtimerendering.com/s2015/AlexEvans_SIGGRAPH-2015-sml.pdf} [Accessed 23-April-2026]}
}

@article{SCHUETZ-2021-PCC,
  title =      "Rendering Point Clouds with Compute Shaders and Vertex Order
               Optimization",
  author =     "Markus Schütz and Bernhard Kerbl and Michael Wimmer",
  year =       "2021",
  month =      jul,
  journal =    "Computer Graphics Forum",
  volume =     "40",
  number =     "4",
  issn =       "1467-8659",
  doi =        "10.1111/cgf.14345",
  booktitle =  "techreport",
  pages =      "12",
  publisher =  "Eurographics Association",
  pages =      "115--126",
  URL =        "https://www.cg.tuwien.ac.at/research/publications/2021/SCHUETZ-2021-PCC/",
}

@article{crane2013robust,
    title={Robust fairing via conformal curvature flow},
    author={Crane, Keenan and Pinkall, Ulrich and Schr{\"o}der, Peter},
    journal={ACM Transactions on Graphics (TOG)},
    volume={32},
    number={4},
    pages={1--10},
    year={2013},
    publisher={ACM New York, NY, USA}
}

@misc{Sponza,
    title        = {Sponza},
    author       = {Marko Dabrovic and Frank Meinl and Crytek and Hans-Kristian Arntzen and Morgan McGuire and Ludicon},
    year         = {2026},
    howpublished = {NVIDIA nvpro-samples},
    note         = {Sponza has undergone several adjustments by different authors over the years. Originally created by Marko Dabrovic, then re-modelled by Frank Meinl at Crytek, Morgan McGuire, Hans-Kristian Arntzen and Ludicon.},
    url          = {https://github.com/ludicon/sponza-gltf}
}

@misc{JapanStatue,
    title        = {Komainu Kobe Ikuta-jinja},
    author       = {Gildas Sidobre, NRHK},
    year         = {2020},
    howpublished = {Distributed by Open Heritage 3D},
    DOI          = {10.26301/1wv3-9775}
}

@misc{molnar1994sorting,
  author       = {Molnar, Steven and Cox, Michael and Ellsworth, David and Fuchs, Henry},
  title        = {A Sorting Classification of Parallel Rendering},
  year         = {1994},
  journal      = {IEEE Computer Graphics and Applications},
  volume       = {14},
  number       = {4},
  pages        = {23--32},
  doi          = {10.1109/38.291528},
  publisher    = {IEEE}
}

@misc{JpegTextures,
      title={Variable-Rate Texture Compression: Real-Time Rendering with JPEG}, 
      author={Elias Kristmann and Michael Wimmer and Markus Schütz},
      year={2025},
      eprint={2510.08166},
      archivePrefix={arXiv},
      primaryClass={cs.GR},
      url={https://arxiv.org/abs/2510.08166}, 
}

@article{RevisitingVertexCache,
    author = {Kerbl, Bernhard and Kenzel, Michael and Ivanchenko, Elena and Schmalstieg, Dieter and Steinberger, Markus},
    title = {Revisiting The Vertex Cache: Understanding and Optimizing Vertex Processing on the modern GPU},
    year = {2018},
    issue_date = {August 2018},
    publisher = {Association for Computing Machinery},
    address = {New York, NY, USA},
    volume = {1},
    number = {2},
    url = {https://doi.org/10.1145/3233302},
    doi = {10.1145/3233302},
    journal = {Proc. ACM Comput. Graph. Interact. Tech.},
    month = aug,
    articleno = {29},
    numpages = {16}
}

@article{Kenzel:2018:OVR,
    author = {Kenzel, Michael and Kerbl, Bernhard and Tatzgern, Wolfgang and Ivanchenko, Elena and Schmalstieg, Dieter and Steinberger, Markus},
    title = {On-the-fly Vertex Reuse for Massively-Parallel Software Geometry Processing},
    journal = {Proc. ACM Comput. Graph. Interact. Tech.},
    issue_date = {August 2018},
    volume = {1},
    number = {2},
    month = aug,
    year = {2018},
    articleno = {28},
    numpages = {17},
    doi = {10.1145/3233303},
    publisher = {ACM},
    address = {New York, NY, USA},
}

@article{10.1145/357332.357335,
    author = {Weghorst, Hank and Hooper, Gary and Greenberg, Donald P.},
    title = {Improved Computational Methods for Ray Tracing},
    year = {1984},
    issue_date = {Jan. 1984},
    publisher = {Association for Computing Machinery},
    address = {New York, NY, USA},
    volume = {3},
    number = {1},
    issn = {0730-0301},
    url = {https://doi.org/10.1145/357332.357335},
    doi = {10.1145/357332.357335},
    journal = {ACM Trans. Graph.},
    month = jan,
    pages = {52–69},
    numpages = {18}
}

@misc{HelperLaneOptimization,
    author = {Bene, Robert and Valasek, Gábor},
    title = {Helper-Lane Optimized Triangulation of Polygons},
    year = {2026},
    publisher = {EUROGRAPHICS},
    note = {short paper}
}

@article{Gnther2013AGP,
    title={A GPGPU-based Pipeline for Accelerated Rendering of Point Clouds},
    author={Christian Günther and Thomas Kanzok and Lars Linsen and Paul Rosenthal},
    journal={J. WSCG},
    year={2013},
    volume={21},
    pages={153-161}
}

@article{ruckert2022adop,
    title={Adop: Approximate differentiable one-pixel point rendering},
    author={R{\"u}ckert, Darius and Franke, Linus and Stamminger, Marc},
    journal={ACM Transactions on Graphics (ToG)},
    volume={41},
    number={4},
    pages={1--14},
    year={2022},
    publisher={ACM New York, NY, USA}
}

@article{NePO_NeuralPointOctrees,
    author = {Lewis, Noah and Rückert, Darius and Stamminger, Marc and Franke, Linus},
    title = {NePO: Neural Point Octrees for Large-Scale Novel View Synthesis},
    journal = {Computer Graphics Forum},
    pages = {e70287},
    doi = {https://doi.org/10.1111/cgf.70287},
    url = {https://onlinelibrary.wiley.com/doi/abs/10.1111/cgf.70287},
    eprint = {https://onlinelibrary.wiley.com/doi/pdf/10.1111/cgf.70287},
}

@Article{kerbl3Dgaussians,
      author       = {Kerbl, Bernhard and Kopanas, Georgios and Leimk{\"u}hler, Thomas and Drettakis, George},
      title        = {3D Gaussian Splatting for Real-Time Radiance Field Rendering},
      journal      = {ACM Transactions on Graphics},
      number       = {4},
      volume       = {42},
      month        = {July},
      year         = {2023},
      url          = {https://repo-sam.inria.fr/fungraph/3d-gaussian-splatting/}
}

@misc{hahlbohm2026fastergs,
    title         = {Faster-GS: Analyzing and Improving Gaussian Splatting Optimization},
    author        = {Florian Hahlbohm and Linus Franke and Martin Eisemann and Marcus Magnor},
    year          = {2026},
    eprint        = {2602.09999},
    archivePrefix = {arXiv},
    primaryClass  = {cs.CV},
    url           = {https://arxiv.org/abs/2602.09999},
}

@misc{RTXMegaGeometry,
    title         = {NVIDIA RTX Mega Geometry Now Available with New Vulkan Samples},
    author        = {Christoph Kubisch and Pyarelal Knowles and Pascal Gautron and Nia Bickford and Jean-Eudes Marvie},
    year          = {2026},
    url           = {https://developer.nvidia.com/blog/nvidia-rtx-mega-geometry-now-available-with-new-vulkan-samples/},
}

@misc{NaniteReyes,
    title         = {Nanite + Reyes},
    author        = {Brian Karis},
    year          = {2026},
    url           = {https://graphicrants.blogspot.com/2026/02/nanite-reyes.html},
}

@article{Piko,
    author = {Patney, Anjul and Tzeng, Stanley and Seitz, Kerry A., Jr. and Owens, John D.},
    title = {Piko: a framework for authoring programmable graphics pipelines},
    year = {2015},
    issue_date = {August 2015},
    publisher = {Association for Computing Machinery},
    address = {New York, NY, USA},
    volume = {34},
    number = {4},
    issn = {0730-0301},
    url = {https://doi.org/10.1145/2766973},
    doi = {10.1145/2766973},
    journal = {ACM Trans. Graph.},
    month = jul,
    articleno = {147},
    numpages = {13}
}

@misc{VisBufferMatGraphs,
    title         = {Visibility Buffer Rendering with Material Graphs},
    author        = {John Hable},
    year          = {2021},
    url           = {http://filmicworlds.com/blog/visibility-buffer-rendering-with-material-graphs/},
}

@mastersthesis{WEBER-2015-PRA1,
  title =      "Micropolygon Rendering on the GPU",
  author =     "Thomas Weber",
  year =       "2015",
  month =      jan,
  address =    "Favoritenstrasse 9-11/E193-02, A-1040 Vienna, Austria",
  school =     "Institute of Computer Graphics and Algorithms, Vienna
               University of Technology ",
  URL =        "https://www.cg.tuwien.ac.at/research/publications/2015/WEBER-2015-PRA1/",
}

@inproceedings{brunhaver2010hardware,
  title={Hardware implementation of micropolygon rasterization with motion and defocus blur.},
  author={Brunhaver, John S and Fatahalian, Kayvon and Hanrahan, Pat},
  booktitle={High Performance Graphics},
  pages={1--9},
  year={2010}
}
\end{document}